\documentclass[a4paper]{article}

\usepackage{natbib}
\usepackage{bbm}
\usepackage{natbib}
\usepackage[all]{xy}\usepackage[latin1]{inputenc}  
\usepackage[dvips]{graphics,graphicx}
\usepackage{amsfonts,amssymb,amsmath,color,mathrsfs, amstext,bbm,dsfont}
\usepackage{amsbsy, amsopn, amscd, amsxtra, amsthm,authblk, enumerate}
\usepackage{upref}
\usepackage{geometry}
\usepackage[displaymath]{lineno}
\usepackage{bm}
\usepackage{dsfont}
\usepackage{todonotes}
\usepackage{mathtools}
\usepackage{float}
\usepackage{comment}

\usepackage{tikz}
\usetikzlibrary{decorations.markings}

\usepackage[all]{xy}
\usepackage[latin1]{inputenc}        
\usepackage[dvips]{graphics,graphicx}
\usepackage{amsfonts,amssymb,amsmath,xcolor,mathrsfs, amstext}
\usepackage{amsbsy, amsopn, amscd, amsxtra, amsthm,authblk, enumerate}
\usepackage{upref}
\usepackage{geometry}
\usepackage[displaymath]{lineno}
\usepackage{float}

\usepackage[colorlinks,
            linkcolor=blue,
            anchorcolor=green,
            citecolor=blue
            ]{hyperref}

\numberwithin{equation}{section}

\def\b{\boldsymbol}

\def\epsilon{\varepsilon}

\newcommand{\ol}{\overline}
\newcommand{\wt}{\widetilde}
\renewcommand{\hat}{\widehat}

\newcommand{\C}{\mathbb{C}}

\newcommand{\bbC}{\mathbb{C}}
\newcommand{\bbD}{\mathbb{D}}
\newcommand{\bbE}{\mathbb{E}}
\newcommand{\E}{\mathbb{E}}

\newcommand{\R}{\mathbb{R}}
\newcommand{\bbR}{\mathbb{R}}
\newcommand{\bbS}{\mathbb{S}}

\newcommand{\bbZ}{\mathbb{Z}}

\newcommand{\cC}{\mathcal{C}}
\newcommand{\cA}{\mathcal{A}}
\newcommand{\cB}{\mathcal{B}}

\newcommand{\cF}{\mathcal{F}}
\newcommand{\cG}{\mathcal{G}}
\newcommand{\cJ}{\mathcal{J}}
\newcommand{\cL}{\mathcal{L}}

\newcommand{\cQ}[2]{\mathcal{Q}\left(#1, #2\right)}

\newcommand{\cS}{\mathcal{S}}
\newcommand{\cT}{\mathcal{T}}

\newcommand{\ld}{\lambda}
\newcommand{\pt}{\partial}

\newcommand{\bD}{\mathbf{D}}

\newcommand{\bL}{\mathbf{L}}
\newcommand{\bbL}{\bar{\mathbf{L}}}

\newcommand{\bS}{\mathbf{S}}

\newcommand{\dd}{\mathrm{d}}
\newcommand{\rv}{\mathrm{v}}
\newcommand{\rw}{\mathrm{w}}

\newcommand{\rD}{\mathrm{D}}
\newcommand{\Mod}{\mathrm{Mod}}
\newcommand{\SLE}{\mathrm{SLE}}
\newcommand{\CLE}{\mathrm{CLE}}

\newcommand{\CR}{\mathrm{CR}}

\newcommand{\ccm}{\b{\mathrm{c}}_{\mathrm{m}}}

\newcommand{\intt}{\mathrm{int}}
\newcommand{\ext}{\mathrm{ext}}

\newcommand{\ind}{\mathds{1}}

\newtheorem{theorem}{Theorem}[section]
\newtheorem{lemma}[theorem]{Lemma}
\newtheorem{proposition}[theorem]{Proposition}
\newtheorem*{proposition*}{Proposition}

\newtheorem*{corollary*}{Corollary}

\newtheorem*{definitions*}{Definitions}

\newtheorem*{example*}{\bf Example}
\theoremstyle{definition}
\newtheorem{remark}[theorem]{Remark}

\numberwithin{equation}{section}
\title{Disk 2-point functions for conformal loop ensemble:\\ a derivation from the SLE loop CFT}
\author[1]{Shengzhi Jin}
\author[1]{Xin Sun}
\author[2]{Baojun Wu}

\affil[1]{Peking University}
\affil[2]{Beijing Institute of Technology}

\date{}

\begin{document}

\maketitle

\begin{abstract}

We give a derivation of the bulk two-point functions of two-dimensional critical loop models in the disk based on conformal loop ensembles (CLE) and the conformal field theory of SLE loop measures developed by Baverez--Jego. Combining local Ward identities with the spectral resolution of the SLE-loop CFT, we express the two-point functions as contour integrals involving known CLE three-point structure constants and Virasoro conformal blocks. 
With a suitable choice of contour, these integrals are absolutely convergent and provide a unified representation of a broad class of correlation functions, valid for both irrational and rational central charges.
For irrational central charges, evaluating these integrals by residues determines the discrete closed-channel spectrum and yields expansions that agree with the analytic bootstrap predictions of Downing--Jacobsen--Nivesvivat--Ribault--Saleur.  

\end{abstract}

\tableofcontents

\section{Introduction}
Two-dimensional critical loop models play a central role in statistical physics, providing a geometric description of many statistical models, such as percolation, the $O(n)$ loop model, and the Fortuin--Kasteleyn (FK) random cluster model. Their critical behavior is expected to be governed by conformal field theory (CFT) \cite{Nienhuis:1984wm,diFrancesco:1987qf}. The probabilistic construction of the Schramm--Loewner evolution (SLE) and conformal loop ensembles (CLE) provides a description of the continuum limit.

Many techniques have made progress in solving these models. The spectrum of the CFT was derived via the Coulomb gas computation of the torus partition function and the transfer-matrix method \cite{diFrancesco:1987qf,DJS2009}. The one-point and three-point structure constants were predicted \cite{DV-connectivity,ijs-cle,ACJNRSW26}, and later justified by the probabilistic method based on SLE and CLE and their coupling theory with Liouville quantum gravity \cite{ssw-radii,ACSW24ImDOZZ,ang2024boundarytouching}. The annulus partition function was also derived via several different methods \cite{SB89,Cardy-2006,ars-annuli}. For higher correlations, the bootstrap method has predicted conformal-block expansion formulae in several cases, including sphere four-point functions \cite{JS-bootstrap,HJJS-bootstrap,Nivesvivat:2023kfp}, disk two-point functions \cite{DJNRS26,DJNS26}, and torus one-point functions \cite{RRJ26Torus}. These bootstrap approaches combine CFT axioms, such as crossing symmetry, with the known spectrum of the theory, to determine the correlations, and thus require a specific treatment for each case. A complete bootstrap solution for general correlations is still hard to achieve.

On the mathematical side, the recent progress on SLE and CLE provides an alternative approach towards conformal bootstrap. The starting point of this approach is the conformal restriction property of SLE loop measures, which describes how their mass changes under local conformal transformations \cite{lsw-restriction,kontsevich2006,zhan-loop-measures}. This property suggests the following formal path-integral interpretation of SLE loop measures \cite{baverez2024cftsleloopmeasures,gordina2024infinitesimal}: 
\[
\SLE_\kappa[\dd\eta] = \exp\left( \frac{\ccm}{24} I^L(\eta) \right) \rD\eta.
\]
Here $\kappa$ is the parameter indexing the loop model, $\ccm=(6-\kappa)(3\kappa-8)/(2\kappa)$ is the central charge, and $\rD\eta$ plays the role of a formal ``uniform measure'' on the loop space. The action $I^L(\eta)$ is the Loewner energy, also known as the universal Liouville action \cite{Wang19Loewner,RW21,TT06}. The Virasoro algebras can then be constructed as geometric operators acting on the loop spaces. Based on this, Baverez and Jego~\cite{baverez2024cftsleloopmeasures} established the basic elements of the CFT of the SLE loop measure, most notably the spectral resolution.

In this paper, we extend the SLE-loop CFT for a single loop to the multiple-loop setting by demonstrating a local Ward identity, and apply the framework to derive bootstrap expansion formulae for bulk two-point functions of critical loop models in the disk. The resulting formulae agree with the corresponding predictions \cite{DJNRS26,DJNS26}. We also recover known annulus partition functions as a sanity check of the method.

\subsection{Disk two-point functions}\label{subsec:intro-result}
We consider critical $O(n)$ loop models in both the dilute and dense regimes, whose conjectural scaling limits are described by $\CLE_\kappa$. We use the parameters \cite{ACJNRSW26}
\begin{equation}\label{eq:intro-parameters}
    \kappa=\frac{4}{b^2}\in\left(\frac{8}{3},8\right),
    \qquad n=-2\cos(\pi b^2),
    \qquad \ccm=1-6(b-b^{-1})^2,
\end{equation}
and the momentum notation
\begin{equation}\label{eq:intro-momenta}
    P_{(r,s)}=\frac{rb-sb^{-1}}{2},
    \qquad \lambda_P=P^2-P_{(1,1)}^2,
    \qquad \Delta_{(r,s)}=\lambda_{P_{(r,s)}}.
\end{equation}
We consider the following observables. Fix two distinct points $0,z\in\bbD$. In the lattice interpretation, loops surrounding only $0$, only $z$, or both points, receive weights $n_1,n_2,n_{\mathrm s}$, respectively, where
\[
    n_j=2\cos(2\pi bP_j),\qquad j=1,2,\mathrm{s}.
\]
Here and throughout the paper, when $\mathrm{s}$ appears in the subscripts, it is viewed as a symbol indicating loops surrounding both points.
Loops surrounding neither point retain the weight $n$.

After suitable local renormalization, the corresponding continuum observable is defined using conformal radii of CLE loops \cite[Lemma 8.1]{ACSW24ImDOZZ}. The conformal radius of a simply connected domain $D$ viewed from an inner point $z\in D$ is defined as $\CR(D;z)=|f'(0)|$, where $f:\bbD\to D$ is the conformal map from the unit disk $\bbD$ to $D$ with $f(0)=z$. For $j=1,2$, let $\eta_j$ be the outermost loop surrounding only the marked point $j$, and let $c_j$ be the logarithm of the conformal radius of the domain bounded by $\eta_j$, viewed from the corresponding marked point. Let $N_{\mathrm{s}}$ count the loops surrounding both points. We consider the following continuum correlation function:
\begin{equation}\label{eq:intro-disk-observable}
    \cG_\bbD(P_1,P_2,P_{\mathrm{s}};z) := \bbE_{\CLE_\bbD}\!\left[ e^{-2\lambda_{P_1}c_1}e^{-2\lambda_{P_2}c_2} \left( \frac{n_{\mathrm{s}}}{n} \right)^{N_{\mathrm{s}}} \right]
\end{equation}

\begin{figure}[htbp]
    \centering
    \includegraphics[width=0.6\textwidth]{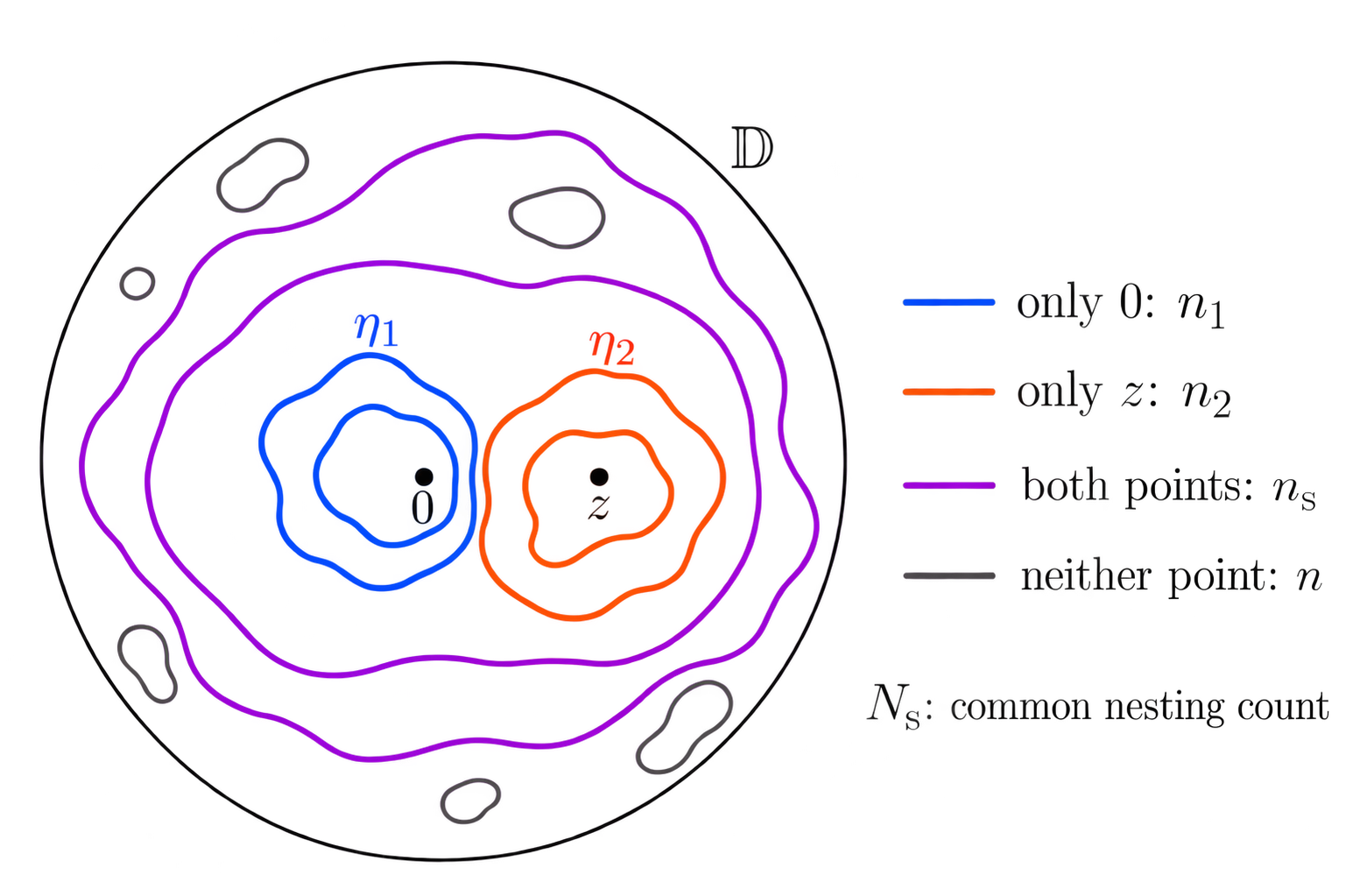}
    \caption{An example of a loop configuration contributing to the two-point function. The weights assigned to the loops according to which marked points they surround are indicated on the right.}
    \label{fig}
\end{figure}

To state our main result, let us abbreviate the chiral Virasoro four-point conformal block by
\[
    \cF_P(x) := \cF^{\ccm}_{\lambda_P} (\lambda_{P_1},\lambda_{P_2},\lambda_{P_2},\lambda_{P_1};x),
\]
whose precise form will be recalled in \eqref{eq:block-definition}.
Let $\hat C$ be the three-point coefficient in the imaginary DOZZ normalization
\begin{equation}
    \hat C_{P_1,P_2}^P = \frac{16P \sin(\pi b^2)\sin(\pi b^{-2})\Gamma_b(2P)\Gamma_b(-2P)}{\prod_{\epsilon_1, \epsilon_2, \epsilon_3 = \pm} \Gamma_b\left(\frac{b + b^{-1}}{2} +\epsilon_1P_1+\epsilon_2P_2 + \epsilon_3 P\right)},
\end{equation}

Our first result is the following bootstrap formula for $\cG_\bbD(P_1,P_2,P_{\mathrm{s}};z)$: 
\begin{equation}\label{eq:intro-integral}
    \boxed{
        \cG_\bbD(P_1,P_2,P_{\mathrm{s}};z) = \mathcal N_1(P_1,P_2) 
        \int_{\mathcal C} \dd\lambda\, \frac{\sin(2\pi b^{-1}P_\lambda) \sin(2\pi bP_\lambda)}{P_\lambda \left(\cos(2\pi bP_\lambda) - \cos(2\pi bP_{\mathrm{s}})\right)} 
        \hat{C}^{P_\lambda}_{P_1,P_2} \cF_{P_\lambda}(|z|^2),
    }
\end{equation}
with properly chosen contour $\mathcal C$ and  the momentum notation given by
\[ P_\lambda = \sqrt{\lambda+\frac{(b-b^{-1})^2}{4}}. \]
The normalization constant $\mathcal N_1(P_1,P_2)$ is independent of $z$, whose explicit expression is recorded in Section~\ref{subsec:disk-2-pt-fct}.
As shown in Section~\ref{subsec:disk-2-pt-fct}, the integrand in \eqref{eq:intro-integral} is a combination of structure constants \eqref{eq:electrical-thickness} \eqref{eq:CLE-jth-1-pt-fct} \eqref{eq:CLE-3pt-fct} which have precise geometric definition through CLE observables.

\begin{figure}[htbp]
    \centering

    \def\dshift{0.75}
    \def\pzerosq{0.04}

    \tikzset{
        contourline/.style={
            thick,
            blue!65!black,
            decoration={
                markings,
                mark=at position .27 with {\arrow{>}},
                mark=at position .73 with {\arrow{>}}
            },
            postaction={decorate}
        }
    }

    \begin{minipage}[t]{0.48\textwidth}
        \centering
        \begin{tikzpicture}[x=.95cm,y=.72cm,>=stealth]
            \pgfmathsetmacro{\vertex}{-\dshift^2-\pzerosq}

            \draw[->] (-1.95,0) -- (3.25,0)
                node[right] {$\mathrm{Re}\lambda$};
            \draw[->] (0,-2.95) -- (0,2.95)
                node[above] {$\mathrm{Im}\lambda$};
            \node[below right] at (0,0) {$0$};

            \draw[contourline]
                plot[domain=-1.8:1.8,samples=90,variable=\u]
                    ({\u*\u+\vertex},{2*\dshift*\u})
                node[above] {$\mathcal C_\delta$};

            \fill[blue!65!black]
                (\vertex,0) circle (1.3pt)
                node[above left,text=black]
                    {$-P_{(1,1)}^2-\delta^2$};
        \end{tikzpicture}
    \end{minipage}
    \hfill
    \begin{minipage}[t]{0.48\textwidth}
        \centering
        \begin{tikzpicture}[x=.95cm,y=.72cm,>=stealth]
            \draw[->] (-2.5,0) -- (2.5,0)
                node[right] {$\mathrm{Re} P$};
            \draw[->] (0,-2.95) -- (0,2.95)
                node[above] {$\mathrm{Im} P$};
            \node[below right] at (0,0) {$0$};

            \draw[contourline]
                (-2.2,\dshift) -- (2.2,\dshift);

            \node[above,blue!65!black] at (1.35,\dshift)
                {$\mathbb R+i\delta$};

            \fill[blue!65!black]
                (0,\dshift) circle (1.3pt)
                node[above left,text=black] {$i\delta$};
        \end{tikzpicture}
    \end{minipage}

    \caption{
        The parabolic contour $\mathcal C_\delta$ in the $\lambda$-plane (left) and the corresponding shifted line $\mathbb R+i\delta$ in the momentum $P$-plane (right), related by $\lambda=P^2-P_{(1,1)}^2$. The arrows indicate the orientation of the contours.
    }
    \label{fig:parabolic-contour}
\end{figure}
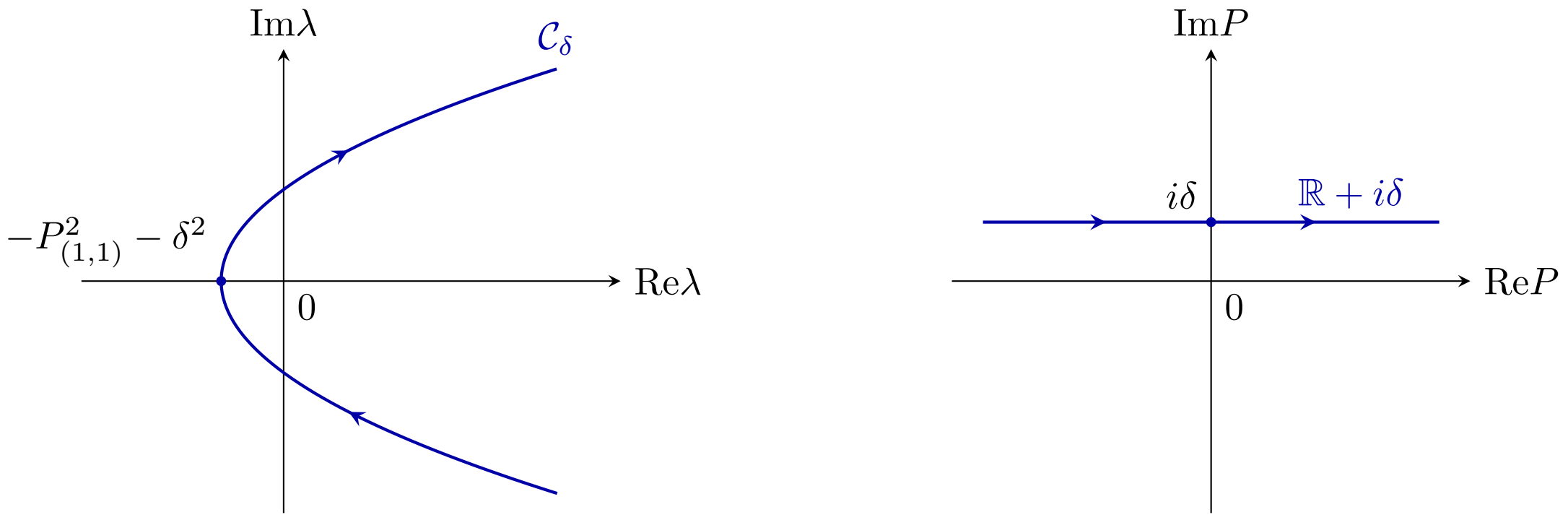

The contour $\mathcal C$ needs to be vertical but bent towards the right to ensure the decay of the integrand.
One natural choice of ${\mathcal C}$ is the parabolic contour
\begin{equation}
 \mathcal C_\delta: \lambda(u) = (u+i\delta)^2 - P_{(1,1)}^2, \qquad u\in\bbR
\end{equation}
oriented in the direction of increasing $u$, for $\delta>0$. 
Figure~\ref{fig:parabolic-contour} illustrates the contour $\mathcal C_\delta$ in the complex $\lambda$-plane. 
This choice was also used in the timelike Liouville theory \cite{Ribault-Santachiara} as the shifted momentum line. A more detailed discussion about the integral contour is given in Section~\ref{subsec:disk-2-pt-fct}.

By formally applying the Residue Theorem, we get a discrete bootstrap expansion for the correlation function:
\begin{equation}\label{eq:intro-disk-expansion}
    \boxed{
        \cG_\bbD(P_1,P_2,P_{\mathrm{s}};z) 
    = \mathcal N_2(P_1,P_2) \cdot \sum_{m\in\bbZ} \sin\left(2\pi b^{-1}P_m\right) \hat{C}^{P_m}_{P_1,P_2} \cF_{P_m}(|z|^2),
    }
\end{equation}
where the summation is over the closed-channel spectrum determined by the intermediate momentum $P_{\mathrm{s}}$: 
\begin{equation}
    P_m=P_{\mathrm{s}}+mb^{-1}, \qquad m\in\bbZ,
\end{equation}
and the expression for the normalization constant $\mathcal N_2(P_1,P_2)$ can be found in Section~\ref{subsec:disk-2-pt-fct}. 
This reproduces the bootstrap expression in \cite{DJNRS26}. For $b\notin\mathbb{R}$, the poles of the integrand in \eqref{eq:intro-integral} form a discrete set, so one may formally obtain the expansion by first applying the residue theorem and then taking the limit as $b$ approaches a real value to get the desired expansion \eqref{eq:intro-disk-expansion}.

Setting the loop weights $n_1$ and $n_2$ to be zero, we recover the bootstrap expression in \cite{DJNRS26,DJNS26} for the two-point connectivity function in the $O(n)$ model and in the FK random-cluster model with free and wired boundary conditions. In the discrete model, these are the probabilities that two marked points belong to the same cluster. Their continuum counterparts are defined using the CLE cluster measure of Miller and Schoug \cite{miller2023existence}. See Section~\ref{subsec:2-pt-connectivity} for more discussion.

Many existing predictions for correlation functions in loop models are expressed as sums over a discrete spectrum. The conformal weights appearing in these expansions have a geometric interpretation and play a central role in approaches based on Temperley--Lieb transfer matrices. However, these expressions have limitations. Such derivation presumes that $b^2$ is irrational to avoid the collision of spectral points and needs a continuation to rational $b^2$. Taking the limit as $b^2$ approaches a rational value is subtle \cite{JaverzatSantachiaraFoda2018}.
Moreover, each summand can also become singular when an internal weight coincides with a Kac pole of the conformal block. 

The contour integral representation offers a unified framework for addressing these issues. The parabolic contour $\mathcal{C}_\delta$ we choose avoids all Kac poles, and the Gaussian decay of the full integrand ensures the absolute convergence of the integral. A detailed discussion about the convergence issue can be found in Section~\ref{subsec:disk-2-pt-fct}. This integral representation provides a uniform expression for a broader class of correlation functions, valid for both generic and degenerate parameter values. By a residue evaluation, we can recover the corresponding discrete summation expansions.

The same framework also applies to more general correlation functions and can reveal additional structures predicted by other approaches. For example, the same strategy yields an integral formula for the disk two-point functions with the boundary interactions introduced in Section~\ref{subsec:JS-annulus}. In this setting, there are additional contributions from the residues of structure constants and conformal blocks labeled by Kac indices. At special parameter values, coalescing poles can produce derivatives of conformal blocks with respect to the internal weight, giving rise to logarithmic blocks. This mechanism can also occur in bulk correlation functions, where one has coincident poles of the holomorphic and antiholomorphic blocks.

\subsection{Derivation from the SLE-loop CFT}

The starting point of our derivation of \eqref{eq:intro-integral} is the one-loop CFT of SLE developed by Baverez and Jego \cite{baverez2024cftsleloopmeasures}. The infinitesimal conformal deformation of a loop defines two commuting Virasoro algebras, and the power of the conformal radius of the loop gives the primary fields. These led to the spectral resolution for functionals on the loop space. Each loop in a CLE configuration is an instance of the SLE loop \cite{ACSW24SLEloop}. The corresponding multiple-loop measures in CLE involve Brownian-loop interaction weights and an additional factor depending on the conformal moduli of the domain bounded by the loops. We demonstrate a local Ward identity involving these weights by a direct geometrical computation. 

To apply the single-loop CFT to compute the disk two-point functions, we cut the loop configuration along a loop separating the marked points from the boundary. Using the domain Markov property of CLE, we can express the correlation functions as the pairing of two functionals of the loop against the SLE measure. The spectral resolution of Baverez and Jego reduces the calculation to computing the pairing of these loop functionals with primary and descendant fields. The primary pairings are given by CLE three-point structure functions, known explicitly from the imaginary DOZZ formula \cite{ACSW24ImDOZZ,ijs-cle}. Ward identities express the descendant pairings as primary terms multiplied by algebraic factors. Summing these algebraic factors over descendants yields the Virasoro conformal blocks.
The spectral resolution therefore leads to an integral along some auxiliary spectral contour involving the known three-point functions and conformal blocks.

Summing over the number of loops surrounding both marked points with their assigned weight gives rise to a factor whose poles depend on the assigned weight. Evaluating the integral \eqref{eq:intro-integral} by residues then gives a discrete sum of conformal blocks and identifies the closed-channel spectrum \eqref{eq:intro-disk-expansion}. The integrand also has Kac poles from the three-point function and the conformal block, but their residues cancel pairwise by Zamolodchikov's recursion relation (Appendix~\ref{appendix:residue-cancellation}).
The integral formula can be derived for all parameter $b$, but the residue-evaluation step requires $b^2$ to be non-rational to avoid the position of the poles to coincide.

So far, the SLE-loop CFT framework based on \cite{baverez2024cftsleloopmeasures} is valid only for the simple loops in the regime of $\SLE_\kappa$ and $\CLE_\kappa$ with $\kappa\in(\frac{8}{3},4)$, which corresponds to the continuum limit of dilute $O(n)$ loop model. However, the structure constants of both the simple-loop case and the non-simple-loop case $\kappa\in(4,8)$ for dense $O(n)$ and FK models have the same analytic expression. This motivates us to propose the same bootstrap formula in the dense
regime. The consistency of the formulae with existing predictions in \cite{DJNRS26} supports this extension.

Before deriving \eqref{eq:intro-disk-expansion}, we first illustrate this method with annulus partition functions, where the primary data are CLE one-point functions and the descendant contributions form Virasoro characters. The same steps of geometric decomposition, spectral integration and residue evaluation recover known $O(n)$ annulus partition functions \cite{SB89,Cardy-2006,ars-annuli} and the FK/Potts expressions with boundary-touching loop reweightings considered here \cite{DJS2009}. These calculations provide sanity checks of our method.

\subsection{Organization of the paper}

Section~\ref{sec:CFT-of-SLE} recalls the SLE-loop CFT and proves the local Ward identity for multiple-loop interactions. Section~\ref{sec:annulus} applies this framework to annulus partition functions. Section~\ref{sec:disk-two-point-functions} derives the disk two-point functions and applies the results to connectivity correlations. Section~\ref{sec:conclusion} summarizes the results and discusses open questions and future directions. Appendix~\ref{appendix:residue-cancellation} gives the residue cancellation needed to identify the discrete spectrum.

\paragraph{Acknowledgement.}
We thank Guillaume Baverez, Jesper Lykke Jacobsen, Rongvoram Nivesvivat and Raoul Santachiara for helpful discussions. The authors were supported by National Key R\&D
Program of China (No. 2023YFA1010700) and the National Natural Science Foundation of China  Grant (No. 12526204).

\paragraph{Statement on AI usage.}
A draft containing the main results was completed before June 2026. The authors used AI for proofreading, language editing and generating figures only. The authors reviewed all suggested changes and take full responsibility for the manuscript.

\section{The CFT of SLE loop measures}\label{sec:CFT-of-SLE}
This section is devoted to the foundation of the SLE loop CFT. We recall the definitions and rigorous results in \cite{baverez2024cftsleloopmeasures} and discuss the extension of spectral resolution in the first subsection, and then demonstrate local Ward identities in the second subsection.

\subsection{The SLE loop Hilbert space and spectral resolution}

\noindent\textbf{Construction of the SLE Hilbert space.}
Consider two marked points on the Riemann sphere $\hat{\bbC}$ (say $0,\infty$), and let $\cJ_{0,\infty}$ be the space of Jordan loops separating $0$ and $\infty$. A loop $\eta\in\cJ_{0,\infty}$ is represented in two ways:
\begin{equation}
  e^{c_\eta} f_\eta : \bbD \to \intt(\eta), \qquad
  e^{-\wt c_\eta} g_\eta : \bbD^* \to \ext(\eta),
\end{equation}
where  $f(0)=0$, $f'(0)=1$, $g(\infty)=\infty$, $g'(\infty)=1$, and
\begin{equation}
    f(z)=z\left(1+\sum_{m\ge1} a_m z^m\right), \qquad
    g(z)=z\left(1+\sum_{m\ge1} b_m z^{-m}\right).
\end{equation}

The SLE loop measure on $\hat\bbC$ can be formally viewed as the path integral
\begin{equation}\label{eq:SLE_path_integral}
    \dd\nu_{\kappa}(\eta) = e^{\frac{\ccm}{24\pi}\bS_1(\eta)} \rD\eta,
\end{equation}
where $\bS_1(\eta)$ is the universal Liouville action given by
\begin{equation}
    \bS_1(\eta) = \int_{\mathbb{D}} \left|\frac{(e^{c_\eta} f_\eta)''}{(e^{c_\eta} f_\eta)'}\right|^{2} |\dd z|^{2} + \int_{\mathbb{D}^{*}} \left|\frac{e^{-\wt c_\eta} g_\eta''}{e^{-\wt c_\eta} g_\eta'}\right|^{2} |\dd z|^{2} + 4\pi\log\left|\frac{(e^{c_\eta} f_\eta)'(0)}{(e^{-\wt c_\eta} g_\eta)'(\infty)}\right|
\end{equation}
for $\eta\in\cJ_{0,\infty}$, and is M\"obius invariant: if $\varphi:\hat\bbC\to\hat\bbC$ is a M\"obius map, then $\bS_1(\eta) = \bS_1(\varphi(\eta))$.

The SLE$_\kappa$ loop measure restricted to $\cJ_{0,\infty}$ is denoted by $\nu$. Its scale invariance gives the factorization
\begin{equation}
    \nu=\frac{\dd c}{\sqrt\pi}\otimes \nu^\#,
\end{equation}
where $\nu^\#$ is the shape measure on normalized conformal maps and is normalized to be a probability measure.

\paragraph{Virasoro operators and the Shapovalov form.}
For a Laurent-polynomial vector field $\rv=v(z)\partial_z\in\bbC(z)\partial_z$, the infinitesimal deformation $\phi_t=\mathrm{id}+t\rv+O(t^2)$ defines complex Lie derivatives $\cL_{\rv},\bar\cL_{\rv}$ by
\begin{equation}
    F(\phi_t\cdot\eta)=F(\eta)+t\cdot\cL_\rv F(\eta)+\bar t\cdot\bar\cL_\rv F(\eta)+o(t).
\end{equation}
Baverez and Jego prove that, on a dense domain of cylindrical smooth functions, these operators are closable and form two commuting Witt representations.  With $\rv_n=-z^{n+1}\partial_z$, this gives $[\cL_n,\cL_m]=(n-m)\cL_{n+m}$ and similarly for the anti-holomorphic copy.

To construct central extension of the Witt algebra, one needs to add the Virasoro correction expressed through Schwarzian one-forms. Define
\begin{equation}
    \vartheta_\eta(\rv) := \frac{1}{2\pi i}\oint_\eta \cS[(e^cf)^{-1}](z)v(z)\dd z, \qquad
    \tilde{\vartheta}_\eta(\rv) := \frac{1}{2\pi i}\oint_\eta \cS[(e^{-\tilde{c}}g)^{-1}](z)v(z)\dd z,
\end{equation}
where $\cS[\cdot]$ is the Schwarzian derivative: $\cS[f] = f'''/f' - \frac{3}{2}(f''/f')^2$. We remark that Schwarzian action is M\"obius invariant: if $\varphi$ is a M\"obius map, then
\begin{equation}\label{eq:Mob-inv-Scharwzian}
    \cS[f] = \cS[\varphi\circ f].
\end{equation}
These two functions give the variation of the action $\bS_1(\eta)$ along the vector field $\rv$:
\begin{equation}
    \cL_\rv(\bS_1(\eta)) = -2\pi\vartheta_\eta(\rv) + 2\pi\tilde{\vartheta}_\eta(\rv).
\end{equation}
The Virasoro operators are 
\begin{equation}
    \bL_\rv:=\cL_\rv-\frac{\ccm}{12}\vartheta(\rv)\mathbf{1}, \qquad \bar\bL_\rv:=\bar\cL_\rv-\frac{\ccm}{12}\overline{\vartheta(\rv)}\mathbf{1}.
\end{equation}
They form two commuting Virasoro algebras with central charge $\ccm$:
\begin{equation}
    [\bL_\rv,\bL_\rw] = \bL_{[\rv,\rw]} + \frac{\ccm}{12}\omega(\rv,\rw)\mathbf{1}, \qquad
    [\bar\bL_\rv,\bar\bL_\rw] = \bar\bL_{[\rv,\rw]} + \frac{\ccm}{12}\overline{\omega(\rv,\rw)}\mathbf{1}, \qquad
    [\bL_\rv,\bar\bL_\rw] = 0,
\end{equation}
where $\omega(\rv,\rw)$ is the Virasoro cocycle:
\begin{equation}
    \omega(\rv,\rw)=\frac{1}{2\pi i}\oint v'''(z)w(z)\dd z.
\end{equation}

We emphasize that the above Virasoro central extension is asymmetric in terms of the two sides of the loop. A key role in constructing the Shapovalov form is played by inversion. Let $\iota(z):=1/\bar z$. The map
\begin{equation}
    \Theta F(\eta) := F(\iota\circ\eta), \qquad \Theta F(c_\eta,f_\eta):=F(\wt c_\eta,\iota\circ g_\eta\circ\iota)
\end{equation}
is a self-adjoint involution of $L^2(\nu)$. Then
\begin{equation}\label{eqdef:Shapovalov_form}
    \cQ{F}{G}:=(F,\Theta G)_{L^2(\nu)}
\end{equation}
is the Shapovalov form, which is non-degenerate but not positive-definite. The integration-by-parts formula by Baverez and Jego implies the Virasoro adjoint relation
\begin{equation}
    \cQ{\bL_{\rv^*}F}{G} = \cQ{F}{\bL_\rv G}, \qquad \cQ{\bar\bL_{\rv^*}F}{G} = \cQ{F}{\bar\bL_\rv G},
\end{equation}
where $\rv^*(z)=z^2\overline{v(1/\bar z)}\,\partial_z$.

\paragraph{Highest weight representations.}
We now turn to the representation theory of the above Virasoro algebras. Let $\cT$ be the set of Young diagrams $\nu=(\nu_1\ge\nu_2\ge\cdots\ge\nu_k)$, and $|\nu|=\sum_{i=1}^k\nu_i$ be its size. Given a Young diagram $\nu$ we denote
\[ \bL_{-\nu} = \bL_{-\nu_k}\cdots\bL_{-\nu_1}, \qquad \bL_\nu=\bL_{\nu_1}\cdots\bL_{\nu_k}, \]
and similarly for the anti-holomorphic copy. 
The actions of the Virasoro operators are explicit in terms of the coordinate functions $(c_\eta,f_\eta)$. In particular, we have for any $\lambda\in\bbC$
\begin{equation}
    \bL_{n>0}(e^{-2\lambda c_\eta}) = \bar\bL_{n>0}(e^{-2\lambda c_\eta})=0, \qquad \bL_0(e^{-2\lambda c_\eta}) = \bar\bL_0(e^{-2\lambda c_\eta}) = \lambda e^{-2\lambda c_\eta},
\end{equation}
which means that $e^{-2\lambda c_\eta}$ is a diagonal primary state with conformal dimension $\lambda$. In addition, Baverez and Jego show the algebraic structure of the descendant states: for any Young diagrams $\nu, \tilde\nu$, 
\[ \bL_{-\nu}\bar\bL_{-\tilde\nu} (e^{-2\ld c_\eta}) = e^{-(2\ld+|\nu|+|\tilde\nu|)c_\eta} \Psi_{\ld,\nu,\tilde\nu}, \]
where $\Psi_{\ld,\nu,\tilde\nu}$ is a polynomial in $\C[(a_m,\bar a_m)_{m\ge1}]$ of total degree $(|\nu|,|\tilde\nu|)$. Here the degree of a monomial $\prod_{j\le n}a_j^{k_j}\bar a_j^{k_j'}$ is defined by $(\sum jk_j, \sum jk_j')$ (which is consistent with the level of the Virasoro operators). In particular, $\Psi_{\ld,\nu,\varnothing}$ is a homogeneous polynomial in $\C[(a_m)_{m\ge1}]$ of degree $|\nu|$. As a result, Baverez and Jego prove that, for non-degenerate weight $\lambda$ (i.e. for those $\lambda$ not in the Kac's table), the set $\{ \Psi_{\ld,\nu,\tilde\nu} \mid \nu,\tilde\nu\in\cT \}$ spans all finite polynomials in $\C[(a_m,\bar a_m)_{m\ge 1}]$ which is dense in $L^2(\nu^\#)$.

\paragraph{Spectral resolution.}
The spectral resolution of $\cQ{\cdot}{\cdot}$ is the computational engine. In this subsection, we focus on the ``sufficiently nice'' functions for which the spectral resolution is rigorously proved by Baverez and Jego, and discuss in the next subsection the extension to other functions involved in solving loop models. We consider the spaces
\begin{equation*}
    \cC:=\cC_c^\infty(\R)\otimes\C[(a_m,\bar{a}_m)_{m\geq1}],
\end{equation*}
where $\cC_c^\infty(\R)$ denotes the space of smooth, compactly supported functions on $\mathbb R$, and 
\begin{equation*}
    \cC_{\mathrm{comp}} := \operatorname{span} \left\{ \eta \mapsto F(c_\eta,f_\eta)\,\chi(c_\eta+\tilde c_\eta) \text{ for some } F\in\cC \text{ and } \chi\in C_c^\infty(\mathbb R)
    \right\}.
\end{equation*}
These spaces are dense subspaces of $L^2(\nu)$. 

We now fix any $\ld_0$, and consider the following set of functions:
\begin{equation*}
    \left\{ \bL_{-\nu}\bar\bL_{-\tilde\nu} (e^{-2\ld c_\eta}) = e^{-(2\ld+|\nu|+|\tilde\nu|)c_\eta} \Psi_{\ld,\nu,\tilde\nu} \mid \nu,\tilde\nu\in\cT, \ld\in\ld_0+i\R \right\}.
\end{equation*}
We will see that this set will play the role of ``orthonormal basis'' with respect to the Shapovalov form.

For Young diagrams $|\nu|=|\nu'|$ we have
\[ \bL_{\nu}\bL_{-\nu'}(e^{-2\lambda c_\eta}) = B_\lambda(\nu,\nu')\cdot e^{-2\lambda c_\eta} \]
where $B_\lambda(\cdot,\cdot)$ is the Gram matrix which is purely determined by the Virasoro commutation relation. The partition function of the Shapovalov form restricted to a highest weight module is given by the moment of the electrical thickness, which can also be viewed as the two-point structure constant of the SLE loop model. To state this precisely, for $\eta\in\cJ_{0,\infty}$, we denote its electrical thickness by
\[ c_{\mathrm{ET}}(\eta):=c_\eta+\tilde{c}_\eta. \]
It also has the following expression:
\begin{equation*}
    c_{\mathrm{ET}}(\eta) = -\frac{1}{2\pi} \int_{\mathbb{D}} \left| \nabla\log\left|\frac{e^{c_\eta}f_\eta(z)}{z}\right| \right|^{2} |\dd z|^{2} - \frac{1}{2\pi} \int_{\mathbb{D}^{*}} \left|\nabla\log\left|\frac{e^{-\tilde c_\eta}g_\eta(z)}{z}\right| \right|^{2} |\dd z|^{2},
\end{equation*}
which implies that $c_{\mathrm{ET}}(\eta)\le0$. The moment of $ c_{\mathrm{ET}}(\eta)$ is computed in \cite{ACSW24SLEloop}: for $\mathrm{Re}(\ld)<\frac{1}{2}(1-\frac{\kappa}{8})$, we have
\begin{equation}\label{eq:electrical-thickness}
    R_2(\ld):= \E_{\nu^\#}[e^{-2\lambda  c_{\mathrm{ET}}(\eta)}] = \frac{\sin\bigl(\pi(1-\kappa/4)\bigr)}{\pi(1-\kappa/4)} \frac{\pi\sqrt{\kappa\lambda+(1-\kappa/4)^2}}{\sin(\pi\sqrt{\kappa\lambda+(1-\kappa/4)^2})},
\end{equation}
or equivalently,
\begin{equation}
    R_2(\ld_P) = \frac{\sin\bigl(\pi(1-b^{-2})\bigr)}{\pi(1-b^{-2})} \frac{2\pi P/b}{\sin(2\pi P/b)}.
\end{equation}

With these two inputs at hand, we can compute the pairing of the ``basis elements''. Formally, the pairing can be stated as:
\begin{equation}
    \cQ{\bL_{-\nu}\bbL_{-\tilde\nu}(e^{-2\lambda c_\eta})}{\bL_{-\nu'}\bbL_{-\tilde\nu'}(e^{-2\ol{\lambda'}c_\eta})} = \delta_{|\nu|=|\nu'|}\delta_{|\tilde\nu|=|\tilde\nu'|} \delta_{\lambda-\lambda'=0} B_\lambda(\nu,\nu') B_\lambda(\tilde\nu,\tilde\nu') R_2(\lambda),
\end{equation}
where the Dirac delta function $\delta_{\lambda-\lambda'=0}$ should be understood in the sense of Fourier transformation in the zero mode $c_\eta$. The term $R_2(\lambda)$ appears in the formula since the Shapovalov form is defined via the involution map $z\mapsto 1/\bar z$. A rigorous statement of this relation is: for any $\mathrm{Re}(\lambda)<\frac{1}{2}(1-\frac{\kappa}{8})$ and any $f\in\cC_c^\infty(\R)$ with Fourier transformation $f(c) = \int_\bbR \frac{\dd p}{\sqrt{\pi}} \hat f(\lambda+ip)e^{-2(\lambda+ip)c}$, we have
\begin{equation}
    \cQ{\bL_{-\nu}\bbL_{-\tilde\nu}(f(c))}{\bL_{-\nu'}\bbL_{-\tilde\nu'}(e^{-2\bar\lambda c_\eta})} = \delta_{|\nu|=|\nu'|}\delta_{|\tilde\nu|=|\tilde\nu'|} \hat f(\lambda) B_{\lambda}(\nu,\nu') B_{\lambda}(\tilde\nu,\tilde\nu') R_2(\lambda).
\end{equation}

As a result, we can perform the spectral decomposition for any function $F\in\cC_\mathrm{comp}$: 
\begin{equation}\label{eq:spectral-decomposition}
    F(\eta) = \int_\bbR \frac{\dd p}{\sqrt{\pi}} \sum_{|\nu|=|\nu'|,|\tilde\nu|=|\tilde\nu'|} \left[\cQ{F}{\bL_{-\nu'}\bbL_{-\tilde\nu'}(e^{-2(\overline{\lambda+ip}) c_\eta})} \frac{B_{\lambda+ip}^{-1}(\nu,\nu') B_{\lambda+ip}^{-1}(\tilde\nu,\tilde\nu')}{R_2(\lambda+ip)}\right] \bL_{-\nu}\bbL_{-\tilde\nu}(e^{-2(\lambda+ip)c_\eta}).
\end{equation}
\begin{remark}
    This integral and summation are proved to be absolutely convergent for the ``nice'' function $F\in\cC_\mathrm{comp}$ \cite{baverez2024cftsleloopmeasures}. In this case, the integral contour can be viewed as $\ld+i\R$ where $\ld$ can be chosen arbitrarily as long as $\mathrm{Re}(\ld)<\frac{1}{2}(1-\frac{\kappa}{8})$. The inverse of the 2-point function $R_2^{-1}$ is an entire function on this domain, while the inverse Gram matrix has poles in Kac's table. However, the corresponding null combinations of descendant fields vanish at the degenerate weight, thus cancel the poles at Kac poles. Thus, the integrand is holomorphic on the domain $\mathrm{Re}(\ld)<\frac{1}{2}(1-\frac{\kappa}{8})$ and suitably deforming the contour does not change the value of the integral.
\end{remark}
\begin{remark}
    A natural question one may ask is whether the spectral decomposition can be extended to a broader class of functions (such as all $L^2(\nu)$) since most of the functions in solving loop models do not lie in the ``nice'' function space $\cC_\mathrm{comp}$. However, this is far from reach if one wants to rigorously prove such mathematical statements. The main obstacle is that the Shapovalov form is indefinite, which means we cannot automatically decompose a $L^2$ function into the basis which is orthogonal with respect to $\cQ{\cdot}{\cdot}$. Another difficulty is that the analytic property (such as growth/decay rate) of the polynomials $\Psi_{\lambda,\nu,\nu'}$ is hard to control, so it is hard to prove the convergence even for some very concrete series. Nevertheless, in this paper we will admit the ``formal'' spectral resolution for some specific functions to derive formulae in the loop models.
\end{remark}

\subsection{Multiple-loop measures and Ward identities}\label{subsec:multi-loop-measure-Ward}
The previous subsection gives the foundation of the single SLE loop CFT. In order to express the CLE correlation functions, it is necessary to consider multiple-loop measures. It is proved by the Liouville quantum gravity techniques that when appropriately choosing several loops in a CLE configuration, the joint law of these loops is closely related to the multiple SLE loop measure \cite{ars-annuli,ACSW24ImDOZZ,BJSW-inprep}. In this subsection, we recall the definition of the multiple SLE loop measure and derive the conformal Ward identity, which will be used later in the spectral decomposition of CLE amplitudes.

Let $n\ge 2$ and $\mathbf{z}=(z_1, \cdots, z_n) \in \hat\C^n$ be $n$ distinct marked points. Define
\[ \mathrm{Loop}_{\mathbf{z}} = \{ [\eta] = (\eta_1, \cdots, \eta_n) \mid \text{$\eta_i$ is a Jordan loop separating $z_i$ from other points, }\eta_i\cap\eta_j=\emptyset \}. \]
For $[\eta] = (\eta_1, \cdots, \eta_n) \in \mathrm{Loop}_{\mathbf{z}}$, define $\intt(\eta_i)$ (resp. $\ext(\eta_i)$) as the connected component of $\hat\C\setminus\eta_i$ containing (resp. not containing) $z_i$.
Let $D_{[\eta]}$ be the $n$-connected component of $\hat\C \setminus (\cup\eta_j)$. We specify the orientation of $\eta_j$ such that it is compatible with the boundary orientation with respect to $D_{[\eta]}$, i.e. such that $D_{[\eta]}$ lies on the left of $\eta_j$.
Let $f_j:\bbD\to\intt(\eta_j)$ and $g_j:\bbD\to\ext(\eta_j)$ be the uniformizing maps. Let
\[ \vartheta_{\eta_j}(\rv) = -\frac{1}{2\pi i}\oint_{\eta_j} \cS[f_j^{-1}](z)v(z)\dd z, \qquad \tilde\vartheta_{\eta_j}(\rv) = \frac{1}{2\pi i}\oint_{\eta_j} \cS[g_j^{-1}](z)v(z)\dd z. \]

Define $\Lambda^*([\eta])=\Lambda^*(\eta_1, \cdots, \eta_n)$ to be the normalized Brownian loop mass, defined inductively by
\[ \Lambda^*(\eta_1, \cdots, \eta_n) = \Lambda^*(\eta_1, \cdots, \eta_{n-1}) + \Lambda^*(\eta_1\cup\cdots\cup\eta_{n-1}, \eta_n), \qquad n\ge 3. \]
The multiple SLE measure is defined by
\begin{equation}
  \dd\nu_{\hat\C,\mathbf{z}}^{(n)}(\eta_1,\ldots,\eta_n) = \ind_{[\eta]\in\mathrm{Loop}_{\mathbf{z}}} \exp\left(\frac{\ccm}{2}\Lambda^*([\eta])\right) \prod_{j=1}^n \dd\nu_{\hat\C}(\eta_j).
\end{equation}
In the following, we will consider meromorphic vector field $\rv$ on $\hat\C$ with poles only at the marked points. Let $\cL_{\rv}^{(j)}$ and $\bar\cL_{\rv}^{(j)}$ be the holomorphic and anti-holomorphic Lie derivative with respect to $\eta_j$.

\begin{proposition}[Ward identity]\label{Thm:WardIdentity}
    For any meromorphic vector field $\rv = v(z)\pt_z$ (with possible poles at $z_1, \cdots, z_n$), we have
    \begin{equation}\label{eq:Ward-identity-Witt}
        \sum_{j=1}^n \cL_\rv^{(j)} \Lambda^*([\eta]) = \sum_{j=1}^n \frac{1}{6}\tilde\vartheta_{\eta_j}(\rv).
    \end{equation}
    Note that this quantity does not depend on the normalization of the uniformizing map $g_j$ thanks to the M\"obius invariance of the Schwarzian derivative \eqref{eq:Mob-inv-Scharwzian}.
\end{proposition}

We now modify the Witt algebras to Virasoro algebras as presented in the previous subsection. For any loop $\eta_j$, we define
\begin{equation}
    \bL_{\rv,\intt}^{(j)}:= \cL_{\rv}^{(j)} - \frac{\ccm}{12}\vartheta_{\eta_j}(\rv), \qquad \bL_{\rv,\ext}^{(j)}:= \cL_{\rv}^{(j)} - \frac{\ccm}{12}\tilde\vartheta_{\eta_j}(\rv).
\end{equation}
Then we have the adjoint relation
\begin{equation}\label{eq:Vir-adjoint}
    \left\langle \bL_{\rv,\intt}^{(j)} F(\eta_j) \mid G(\eta_j)\right\rangle_{\dd\nu_{\hat\C}(\eta_j)} = - \left\langle F(\eta_j) \mid \bL_{\rv,\ext}^{(j)} G(\eta_j)\right\rangle_{\dd\nu_{\hat\C}(\eta_j)}.
\end{equation}
The above Ward identity for the n-loop amplitude can be rewritten as
\begin{equation}
    \sum_{j=1}^n \bL_{\rv,\ext}^{(j)} e^{\frac{\ccm}{2}\Lambda^*([\eta])} = 0.
\end{equation}
Thus, we have the following formulation of the Ward identity:
\begin{equation}
    \left\langle \bL_{\rv,\intt}^{(1)} F([\eta]) \right\rangle_{\dd\nu_{\hat\C,\mathbf{z}}^{(n)}([\eta])} = -\sum_{j=2}^n \left\langle \bL_{\rv,\intt}^{(j)} F([\eta]) \right\rangle_{\dd\nu_{\hat\C,\mathbf{z}}^{(n)}([\eta])}
\end{equation}

The rest of the section is devoted to the proof of Proposition \ref{Thm:WardIdentity}. We follow the approach in \cite[Section 4]{baverez2024cftsleloopmeasures}, where they compute the variation of the normalized Brownian loop measure for two loops.

We first prove the following lemma, which computes the Brownian bubble measure for a multi-connected domain. Let $\Psi:\bbD\to D_{[\eta]}$ be the holomorphic universal covering, and let $G$ be the fundamental group
\[ G:= \{ g\in\mathrm{Aut}(\bbD) \mid \Psi\circ g=\Psi \}. \]
For any $z\in\eta_j$, we denote by $n_z\in\bbS^1$ the outer normal vector at $z$, which implies that $in_z$ is the unit tangent vector along $\eta_j$ at $z$.

\begin{lemma}\label{lem:bubble_multi_connected}
    For any $z \in \eta_j$, we have
    \[ \mu_{\ext(\eta_j), z}^{\mathrm{bub}}(\mathrm{hit}\;\cup_{i\ne j}\eta_i) = -\frac{n_z^2}{6} \cS[g_j^{-1}](z) + \frac{n_z^2}{6} \cS[\Psi^{-1}](z) + n_z^2 \sum_{g\in G\setminus\{\mathrm{id}\}} \frac{(\Psi'(\tilde z))^{-1} ((\Psi\circ g^{-1})'(g(\tilde z)))^{-1}}{(\tilde z-g(\tilde z))^2} \]
\end{lemma}
\begin{proof}
    We start by writing
    \begin{equation}\label{eq:proof_of_Ward}
    \begin{split}
        \mu_{\ext(\eta_j), z}^{\mathrm{bub}}(\mathrm{hit}\;\cup_{i\ne j}\eta_i) &= \pi\lim_{w\to z}\mu_{\ext(\eta_j), z, w}^{\mathrm{exc}}(\mathrm{hit}\;\cup_{i\ne j}\eta_i) \\
        &= \pi\lim_{w\to z} \left( H_{\ext(\eta_j)}(z,w) - H_{D_{[\eta]}}(z,w) \right),
    \end{split}
    \end{equation}
    where $\mu_{\Omega,x,y}^{\mathrm{exc}}$ is the Brownian excursion measure in the domain $\Omega$ from $x$ to $y$ and its total mass is the boundary Poisson kernel $H_\Omega(x,y)$. Recall that the boundary Poisson kernel on the unit disk is $H_\bbD(x,y)=\frac{1}{\pi|x-y|^2} = -\frac{xy}{\pi(x-y)^2}$. 
    
    We first consider the first term in \eqref{eq:proof_of_Ward}. Using the covariant property of the boundary Poisson kernel and the relation $|(g_j^{-1})'(z)| = (g_j^{-1})'(z) \frac{n_z}{g_j^{-1}(z)}$, we have
    \[ \pi H_{\ext(\eta_j)}(z,w) = \pi |(g_j^{-1})'(z)| |(g_j^{-1})'(w)| H_\bbD(g_j^{-1}(z), g_j^{-1}(w)) = -n_zn_w\frac{(g_j^{-1})'(z) (g_j^{-1})'(w)}{(g_j^{-1}(z)-g_j^{-1}(w))^2}. \]
    An elementary Taylor expansion gives
    \[ \lim_{w\to z} \frac{(g_j^{-1})'(z) (g_j^{-1})'(w)}{(g_j^{-1}(z)-g_j^{-1}(w))^2} - \frac{1}{(z-w)^2} = \frac{1}{6}\cS[g_j^{-1}](z), \]
    and thus
    \[ \lim_{w\to z} \pi H_{\ext(\eta_j)}(z,w) + n_zn_w \frac{1}{(z-w)^2} = -\frac{n_z^2}{6} \cS[g_j^{-1}](z) \]
    
    For the second term $H_{D_{[\eta]}}(z,w)$ in \eqref{eq:proof_of_Ward}, we can compute it using its universal covering $\Psi:\bbD\to D_{[\eta]}$. Let $\tilde z\in\bbS^1$ be any preimage of $z$ under the covering map $\Psi$, and $\tilde w$ be the preimage of $w$ such that $\tilde w\to\tilde z$ as $w\to z$. Then we have
    \[ \pi H_{D_{[\eta]}}(z,w) = \pi \sum_{g\in G} |\Psi'(\tilde z)|^{-1} |(\Psi\circ g^{-1})'(g(\tilde w))|^{-1} H_\bbD(\tilde z, g(\tilde w)). \]
    For the summand with $g=\mathrm{id}$, a similar computation as above gives
    \begin{align*}
        \lim_{w\to z} \pi |\Psi'(\tilde z)|^{-1} |\Psi'(\tilde w)|^{-1} H_\bbD(\tilde z, \tilde w) + n_zn_w \frac{1}{(z-w)^2}
        = -\frac{n_z^2}{6} \cS[\Psi^{-1}](z).
    \end{align*}
    where we use $\Psi^{-1}(z)$ with a slight abuse of notation to denote any local inverse of $\Psi$ and there is no ambiguity thanks to the M\"obius invariance of the Schwarzian derivative, i.e. $\cS[\Psi^{-1}](z) = \cS[(\Psi\circ g)^{-1}](z)$ for any $g\in G$.
    
    For the remaining terms in the summation, there is no singularity in the limit and we have:
    \begin{align*}
        & \lim_{w\to z} \pi \sum_{g\in G\setminus\{\mathrm{id}\}} |\Psi'(\tilde z)|^{-1} |(\Psi\circ g^{-1})'(g(\tilde w))|^{-1} H_\bbD(\tilde z, g(\tilde w)) \\
        =& \lim_{w\to z} \sum_{g\in G\setminus\{\mathrm{id}\}} -n_zn_w \frac{(\Psi'(\tilde z))^{-1} ((\Psi\circ g^{-1})'(g(\tilde w)))^{-1}}{(\tilde z-g(\tilde w))^2}
        = -n_z^2 \sum_{g\in G\setminus\{\mathrm{id}\}} \frac{(\Psi'(\tilde z))^{-1} ((\Psi\circ g^{-1})'(g(\tilde z)))^{-1}}{(\tilde z-g(\tilde z))^2}.
    \end{align*}
    This summation is finite and uniformly bounded for $z\in\eta_j$ since it is the total mass of non-contractible Brownian excursions in $D_{[\eta]}$ rooted at $z$.
    
    Adding together, we get the result.

\end{proof}

For $z\in\overline{D_{[\eta]}}$, write
\[ h(z) := \frac{1}{6} \cS[\Psi^{-1}](z) +  \sum_{g\in G\setminus\{\mathrm{id}\}} \frac{(\Psi'(\tilde z))^{-1} ((\Psi\circ g^{-1})'(g(\tilde z)))^{-1}}{(\tilde z-g(\tilde z))^2}. \]
We will show later in the proof of Proposition \ref{Thm:WardIdentity} that $h$ is well-defined and is holomorphic in $D_{[\eta]}$.

\begin{lemma}
    For any holomorphic vector flow $\rv = v(z)\pt_z$ (with possible poles at $z_1, \cdots, z_n$), we have
    \begin{equation}\label{eq:Lie_j}
        \cL_\rv^{(j)} \Lambda^*([\eta]) = \frac{1}{6}(\cS[g_j^{-1}],\rv) - \frac{1}{2\pi i}\oint_{\eta_j} h(z) v(z)\dd z
    \end{equation}
\end{lemma}
\begin{proof}
    By the linearity of Lie derivative, we can assume for convenience that $\rv$ is Markovian in the sense that $\mathrm{Re}(-v(z)/n_z)>0$. Following the same strategy as in \cite[Lemma 4.7]{baverez2024cftsleloopmeasures}, we have
    \begin{align*}
        &\lim_{t\to0, t>0} \frac{1}{t} \left( \Lambda^*(\eta_1, \cdots, e^{t\rv}(\eta_j), \cdots, \eta_n) - \Lambda^*(\eta_1, \cdots, \eta_n) \right) \\
        =& -\frac{1}{\pi}\oint_{\eta_j} |\dd z| \mathrm{Re}\left(\frac{v(z)}{n_z}\right) \mu_{\ext(\eta_j),z}^{\mathrm{bub}}(\mathrm{hit}\;\cup_{i\ne j}\eta_i) 
        = \mathrm{Re}\left[ -\frac{1}{\pi} \oint_{\eta_j} \frac{\dd z}{in_z^2} v(z) \mu_{\ext(\eta_j),z}^{\mathrm{bub}}(\mathrm{hit}\;\cup_{i\ne j}\eta_i) \right]\\
        =&\; \mathrm{Re}\left[ \frac{1}{3}(\cS[g_j^{-1}],\rv) - \frac{1}{\pi i}\oint_{\eta_j} h(z) v(z)\dd z \right].
    \end{align*}
    The result then follows from the definition of $\cL_{\rv}^{(j)}$.
\end{proof}

\begin{proof}[Proof of Proposition \ref{Thm:WardIdentity}]
    For each $g\in G\setminus\{\mathrm{id}\}$, write
    \[ h_g(z) := \frac{(\Psi'(\tilde z))^{-1} ((\Psi\circ g^{-1})'(g(\tilde z)))^{-1}}{(\tilde z-g(\tilde z))^2}. \]
    In the proof of Lemma \ref{lem:bubble_multi_connected} we have shown that for $z\in\pt D_{[\eta]}$ we have $-n_z^2h_g(z)\ge 0$, and $\sum_{g\in G\setminus\{\mathrm{id}\}} |h_g(z)|$ is the total mass of non-contractible Brownian excursions and thus is uniformly bounded. Note that for each $g$, $h_g(z)$ is a holomorphic function on the domain of a chosen local inverse branch of $\Psi$. The series converges locally normally. Changing the lift $\tilde z$ only conjugates the index $g$, hence permutes this absolutely convergent sum.
    Thus $\sum_{g\in G\setminus\{\mathrm{id}\}} h_g(z)$ is a single-valued bounded holomorphic function in $D_{[\eta]}$ and is continuous up to boundary, and so is $h(z)$.

    We now obtain from the holomorphicity of $h(z)$ that
    \[ \sum_{j=1}^n \oint_{\eta_j} h(z)v(z) \dd z =0. \]
    Summing up \eqref{eq:Lie_j} over $j=1,\cdots,n$ concludes the proof.
\end{proof}

\section{Annulus partition functions}\label{sec:annulus}
\subsection{Boundary states and spectral resolution}
In this subsection, we first introduce the boundary state as a function of a SLE loop, and then study its spectral resolution. To illustrate the basic ideas, we will consider the annulus SLE partition function which is the easiest example to see the algebraic structure while lacks a direct physical interpretation. In the later subsections, we will follow the same strategy to compute annulus partition functions in several loop models, which are predicted previously by other physical methods and are proved by the rigorous Liouville quantum gravity coupling technique.

Let $A_q=\bbD\setminus q\bbD$, $0<q<1$, and write $q=e^{-2\pi\tau}$ with $\tau>0$. Let $\SLE_{A_q,\kappa}$ denote the SLE loop measure in $A_q$ restricted to loops separating the two boundary components:
\begin{equation}
    \SLE_{A_q,\kappa}[\dd\eta] = \exp\left(\frac{\ccm}{2}\Lambda^*(\eta,\bbS^1\cup q\bbS^1)\right) \ind_{\eta\subset A_q} \dd\nu_{\hat\C}(\eta).
\end{equation}
By a simple inclusion-exclusion principle, the normalized loop mass can be written as
\begin{equation*}
    \Lambda^*(\eta,\bbS^1\cup q\bbS^1) = \Lambda^*(\eta,\bbS^1) + \Lambda^*(\eta,q\bbS^1)-\Lambda^*(\bbS^1,q\bbS^1).
\end{equation*}
Define two boundary states to be
\begin{equation}\label{eq:SLE-boundary-state}
    \cB_1(\eta) = \exp\left(\frac{\ccm}{2}\Lambda^*(\eta,\bbS^1)\right) \ind_{\eta\subset\bbD}, \qquad 
    \cB_2(\eta) = \exp\left(\frac{\ccm}{2}\Lambda^*(\eta,q\bbS^1)\right) \ind_{\eta\subset q\bbD^*},
\end{equation}
then the annulus SLE partition function can be viewed as the pairing of these two states:
\begin{equation}\label{eq:decomp-SLE-annulus-p-f}
    |\SLE_{A_q,\kappa}| = \exp\left(-\frac{\ccm}{2}\Lambda^*(\bbS^1,q\bbS^1)\right) \langle \cB_1(\eta)\mid\cB_2(\eta)\rangle_{\dd\nu_{\hat\C}(\eta)}.
\end{equation}

We now study the algebraic property of the boundary states. For $n\ge 0$, the real-parameter vector flows generated by $\rv_n-q^{2n}\rv_{-n}$ and $i(\rv_n+q^{2n}\rv_{-n})$ preserve the boundary $q\bbS^1$. Thus the Ward identity \eqref{eq:Ward-identity-Witt} (applying to the pair $\eta,\bbS^1$) gives:
\begin{equation}\label{eq:boundary-state-contition-real}
    \mathrm{Re}\left[ (\bL_n-q^{2n}\bL_{-n})\cB_2(\eta) \right]=0, \qquad \mathrm{Im}\left[(\bL_n+q^{2n}\bL_{-n})\cB_2(\eta) \right]=0,
\end{equation}
and thus
\begin{equation}\label{eq:boundary-state-condition}
    \bL_{n}\cB_2(\eta) = q^{2n}\bbL_{-n}\cB_2(\eta).
\end{equation}
A similar relation holds for $\cB_1(\eta)$, with $q$ replaced by $1$ and the Virasoro operators replaced by their conjugation under the involution $\Theta$.

We formally extend the spectral decomposition \eqref{eq:spectral-decomposition} to the boundary states where the coefficients can be determined by the algebraic relation. For any $\nu',\tilde\nu'\in\cT$, we have
\begin{align*}
    &\; \cQ{\cB_2}{\bL_{-\nu'}\bbL_{-\tilde\nu'}(e^{-2\bar\lambda c_\eta})} = \cQ{\bL_{\nu'}\cB_2}{\bbL_{-\tilde\nu'}(e^{-2\bar\lambda c_\eta})} \\
    =&\; \cQ{q^{2|\nu'|}\bbL_{-\nu'}\cB_2}{\bbL_{-\tilde\nu'}(e^{-2\bar\lambda c_\eta})} = q^{2|\nu'|}\cQ{\cB_2}{\bbL_{\nu'}\bbL_{-\tilde\nu'}(e^{-2\bar\lambda c_\eta})}.
\end{align*}
If $|\nu'|>|\tilde\nu'|$, then $\bbL_{\nu'}\bbL_{-\tilde\nu'}(e^{-2\bar\lambda c_\eta})=0$ since $e^{-2\bar\lambda c_\eta}$ is a primary state and the pairing vanishes. Similarly, it vanishes if $|\nu'|<|\tilde\nu'|$. The only remaining case is $|\nu'|=|\tilde\nu'|$ which gives
\begin{equation}
    \cQ{\cB_2}{\bL_{-\nu'}\bbL_{-\tilde\nu'}(e^{-2\bar\lambda c_\eta})} = q^{2|\tilde\nu'|}B_\lambda(\nu',\tilde\nu')\cQ{\cB_2}{e^{-2\bar\lambda c_\eta}}.
\end{equation}
The pairing of $\cB_2$ with primary state is given by the disk one-point function: by the definition of disk SLE measure $\SLE_{q\bbD^*}[\dd\eta] = \cB_2(\eta)\dd\nu_{\hat\C}(\eta)$, we have
\[ \cQ{\cB_2}{e^{-2\bar\lambda c_\eta}} = \SLE_{q\bbD^*}[e^{-2\lambda\tilde c_\eta}] = q^{2\lambda} S_1(\lambda), \]
where $S_1$ is the SLE disk one-point function:
\[ S_1(\lambda) = \SLE_{\bbD}[e^{-2\lambda c_\eta}\ind_{0\in\intt(\eta)}]. \]
Thus, by \eqref{eq:spectral-decomposition} we have the spectral decomposition:
\begin{equation}\label{eq:decomp-cB_2}
\begin{split}
    \cB_2(\eta) =& \int_\bbR \frac{\dd p}{\sqrt{\pi}} \sum_{|\nu|=|\nu'|=|\tilde\nu|=|\tilde\nu'|} \left[q^{2(\lambda+ip)+2|\tilde\nu'|} S_1(\lambda+ip) B_{(\lambda+ip)}(\nu',\tilde\nu') \frac{B_{\lambda+ip}^{-1}(\nu,\nu') B_{\lambda+ip}^{-1}(\tilde\nu,\tilde\nu')}{R_2(\lambda+ip)}\right] \bL_{-\nu}\bbL_{-\tilde\nu}(e^{-2(\lambda+ip)c_\eta}) \\
    =& \int_\bbR \frac{\dd p}{\sqrt{\pi}} \sum_{|\nu|=|\tilde\nu|} q^{2(\lambda+ip)+2|\nu|} B_{\lambda+ip}^{-1}(\tilde\nu,\nu) \frac{S_1(\lambda+ip)}{R_2(\lambda+ip)} \bL_{-\nu}\bbL_{-\tilde\nu}(e^{-2(\lambda+ip)c_\eta}).
\end{split}
\end{equation}
Pairing the above expression with $\cB_1(\eta)$, we have
\begin{equation}
\begin{split}
    \left\langle \cB_1(\eta) \mid \cB_2(\eta) \right\rangle =& \int_\bbR \frac{\dd p}{\sqrt{\pi}} \sum_{|\nu|=|\tilde\nu|} q^{2(\lambda+ip)+2|\nu|} B_{\lambda+ip}^{-1}(\tilde\nu,\nu) \frac{S_1(\lambda+ip)}{R_2(\lambda+ip)} \left\langle \cB_1(\eta) \mid \bL_{-\nu}\bbL_{-\tilde\nu}(e^{-2(\lambda+ip)c_\eta}) \right\rangle \\
    =& \int_\bbR \frac{\dd p}{\sqrt{\pi}} \sum_{|\nu|=|\tilde\nu|} q^{2(\lambda+ip)+2|\nu|} B_{\lambda+ip}^{-1}(\tilde\nu,\nu) \frac{S_1(\lambda+ip)}{R_2(\lambda+ip)} B_{\lambda+ip}(\tilde\nu,\nu) S_1(\lambda+ip) \\
    =& \int_\bbR \frac{\dd p}{\sqrt{\pi}} \frac{S_1(\lambda+ip)^2}{R_2(\lambda+ip)} q^{2(\lambda+ip)} \left[\sum_{n\ge0} |\cT_n|q^{2n}\right]
    = \int_\bbR \frac{\dd p}{\sqrt{\pi}} \frac{S_1(\lambda+ip)^2}{R_2(\lambda+ip)} \frac{q^{2(\lambda+ip+\frac{1}{24})}}{\eta(\frac{\log q}{\pi i})},
\end{split}
\end{equation}
where $\eta$ is the Dedekind eta function.

According to \cite[Proposition 4.2]{LM25-2loop}, the pre-factor in \eqref{eq:decomp-SLE-annulus-p-f} is given by 
\[ \Lambda^*(\bbS^1, e^{-2\pi\tau}\bbS^1) = \log \left(\eta(2i\tau)^{-2} \tau^{-1} e^{-\pi\tau/3} \right) + const.\]
Thus, we derive the annulus SLE partition function:
\begin{equation}\label{eq:SLE-annulus-p-f}
\begin{split}
    |\SLE_{A_q,\kappa}| =&\; C\cdot \eta(2i\tau)^{\ccm}\tau^{\ccm/2} \int_\bbR \frac{\dd p}{\sqrt{\pi}} \frac{S_1(\lambda+ip)^2}{R_2(\lambda+ip)} \frac{e^{-4(\lambda+ip+\frac{1-\ccm}{24})\pi\tau}}{\eta(2i\tau)} \\
    =&\; C\cdot \eta(2i\tau)^{\ccm}\tau^{\ccm/2} \int P\dd P\frac{S_1(\lambda_P)^2}{R_2(\lambda_P)} \frac{e^{-4P^2\pi\tau}}{\eta(2i\tau)},
\end{split}
\end{equation}
where, in the last line, the integral contour is a branch of $\{ P\in\C\mid \mathrm{Re}(\lambda_P)=\lambda_0 \}$ for some $\lambda_0$, and $\frac{e^{-4P^2\pi\tau}}{\eta(2i\tau)}$ is the Virasoro character. Here and in the rest of the Section, $C$ is some constant independent of the modulus $\tau$ of the annulus.

Although we will not use the result in this paper, we remark that the SLE one-point function $S_1(\lambda_P)$ can be derived from the Liouville quantum gravity coupling technique: for $\mathrm{Re(\lambda_P)}<0$, we have
\begin{equation*}
    S_1(\lambda_P) = C\cdot R_2(\lambda_P) \int_0^\infty e^{4\pi P^2\tau} \frac{\dd\tau}{\tau^{\ccm/2} \eta(2i\tau)^{\ccm-1}}.
\end{equation*}

Our analytic derivation partly follows the derivation in \cite{Cardy1989BCFT}. With the algebraic condition \eqref{eq:boundary-state-condition} on the boundary state, in \eqref{eq:decomp-cB_2} we formally decompose the state as an integral of Ishibashi states with weights $\lambda_P$, and the coefficients are determined by the one-point functions. However, we do not identify the spectrum of the theory. What we do is formally performing Fourier transformation on the zero-mode $c_\eta$. The integral contour in \eqref{eq:SLE-annulus-p-f} can be chosen arbitrarily if the Residue Theorem can be applied, and does not yet have a physical meaning. We regard this as an intermediate step in deriving the partition functions in loop models, and we will see in the following subsections that we can actually identify the spectrum of the theory from this formal integral.

\subsection{Dilute O(n) loop models on the annulus}\label{subsection:annulus-O(n)-p-f}

The scaling limit of dilute $O(n)$ loop model is conjectured to be simple conformal loop ensemble $\CLE_\kappa$ with the parameter related as follows:
\[ \kappa = \frac{4}{b^2}\in(8/3,4], \qquad n = -2\cos(\pi b^2) = 2\cos(2\pi bP_{(1,1)})\in(0,2]. \]
The annulus CLE is defined in \cite{ars-annuli}, which we recall in the following. Consider a disk $\CLE_\bbD$ and let $\Gamma$ be its configuration. Let $(\eta_k)_{k\ge1}$ be the sequence of loops in $\Gamma$ surrounding the origin, ordered such that $\eta_{k+1}$ is surrounded by $\eta_k$.
We choose $\eta$ by counting measure of $\{\eta_k\}$, and let $A_\eta$ be the annulus bounded by $\eta$ and $\partial\bbD$. We define the annulus $\CLE_{\kappa}^\tau$ to be the conditional law of the configuration $\Gamma$ inside $A_\eta$ conditioning on $\Mod(A_\eta) = \tau$. The conformal invariance of SLE guarantees that it is well-defined. 

The partition function of the $O(n)$ loop model can be regarded as the distribution of the random modulus $\Mod(\eta)$. Furthermore, we can reweight all non-contractible loops in the annulus CLE configuration by $\frac{n'}{n}$ (or equivalently, reweight each configuration by $\left(\frac{n'}{n}\right)^{k-1}$ if $\eta=\eta_k$) and get the partition function of the annulus $O(n,n')$ model.

These partition functions were derived in \cite{SB89} via the quantum group method, derived in \cite{Cardy-2006} via the Coulomb gas method, and derived in \cite{ars-annuli} via rigorous Liouville quantum gravity coupling method. We now use our new analytic framework to derive them.

For $k\ge 1$, let $Z_k(\tau)$ be the density for the random modulus $\Mod(A_{\eta_k})$. Let $Z_{\SLE}(\tau)$ be the density for the random modulus $\Mod(\tilde\eta)$ of $\tilde\eta$ sampled from the disk SLE measure $\SLE_{\bbD}[\dd\tilde\eta] \ind_{\tilde\eta\text{ surrounds 0}}$. A key input is that for each $k\ge1$, the law of $\eta_k$ differs from the disk SLE measure only by a factor of the conformal modulus, i.e. it has the same distribution as
\[ \frac{Z_k(\tau_\eta)}{Z_{\SLE}(\tau_\eta)} \SLE_{\bbD}[\dd\eta] \ind_{\eta\text{ surrounds 0}} = \frac{Z_k(\tau_\eta)}{Z_{\SLE}(\tau_\eta)} e^{\frac{\ccm}{2}\Lambda^*(\eta,\bbS^1)} \dd\nu_{\hat\C}(\eta). \]
This is implicitly implied by the LQG coupling technique in \cite{ars-annuli}.

The $O(n,n')$ partition function for an annulus of modulus $\tau$ can be written as
\[ Z(\tau;n') = \sum_{k\ge1} \left(\frac{n'}{n}\right)^{k-1} Z_k(\tau). \]
We first compute $Z_k(\tau)$ for $k\ge2$ and then discuss its extension to $k=1$. For any pair $(\eta,\eta')$ that bounds an annulus, we denote by $\tau_{\eta,\eta'}$ the annulus modulus, and write $\tau_\eta = \tau_{\eta,\bbS^1}$ for short. By the domain Markov property of CLE, the law of $\eta_k$ is the same as $\eta_{k-1}'$ sampled as follows: we first sample $\eta_1$ from $\CLE_\bbD$ as above, and then sample an independent CLE configuration $\Gamma'$ inside $\eta_1$ and let $\eta_{k-1}'$ be the $(k-1)$-th outermost loop in $\Gamma'$ surrounding the origin. Thus, for any test function $f$ we have
\begin{equation}
\begin{split}
    &\; \CLE_{\bbD}[f(\tau_{\eta_k})] = \int f(\tau_\eta) \frac{Z_k(\tau_\eta)}{Z_{\SLE}(\tau_\eta)} e^{\frac{\ccm}{2}\Lambda^*(\eta,\bbS^1)} \dd\nu_{\hat\C}(\eta) \\
    =&\; \int \CLE_{\bbD}[\dd\eta_1] \left( \int \CLE_{\intt(\eta_1)}[\dd\eta_{k-1}'] f(\tau_{\eta_{k-1}'}) \right) \\
    =&\; \int \frac{Z_1(\tau_{\eta_1})}{Z_{\SLE}(\tau_{\eta_1})} e^{\frac{\ccm}{2}\Lambda^*(\eta_1,\bbS^1)} \dd\nu_{\hat\C}(\eta_1) \left( \int\frac{Z_{k-1}(\tau_{\eta_1,\eta})}{Z_{\SLE}(\tau_{\eta_1,\eta})} e^{\frac{\ccm}{2}\Lambda^*(\eta_1,\eta)} f(\tau_\eta) \ind_{\eta_1\subset A_\eta} \dd\nu_{\hat\C}(\eta) \right) \\
    =&\; \int e^{\frac{\ccm}{2}\Lambda^*(\eta,\bbS^1)} f(\tau_\eta)\dd\nu_{\hat\C}(\eta) \left( \int\frac{Z_1(\tau_{\eta_1})}{Z_{\SLE}(\tau_{\eta_1})} \frac{Z_{k-1}(\tau_{\eta_1,\eta})}{Z_{\SLE}(\tau_{\eta_1,\eta})} e^{\frac{\ccm}{2}\Lambda^*(\eta_1,\bbS^1\cup\eta)} \ind_{\eta_1\subset A_\eta} \dd\nu_{\hat\C}(\eta_1)\right).
\end{split}
\end{equation}
Note that the measure $e^{\frac{\ccm}{2}\Lambda^*(\eta_1,\bbS^1\cup\eta)} \dd\nu_{\hat\C}(\eta_1)$ is the annulus SLE measure in the annulus $A_\eta$. By the conformal invariance of the annulus SLE measure, the term in the bracket in the last line depends only on $\tau_\eta$. Thus the above identity gives
\begin{equation}
    \frac{Z_k(\tau_\eta)}{Z_{\SLE}(\tau_\eta)} = \int\frac{Z_1(\tau_{\eta_1})}{Z_{\SLE}(\tau_{\eta_1})} \frac{Z_{k-1}(\tau_{\eta_1,\eta})}{Z_{\SLE}(\tau_{\eta_1,\eta})} e^{\frac{\ccm}{2}\Lambda^*(\eta_1,\bbS^1\cup\eta)} \ind_{\eta_1\subset A_\eta} \dd\nu_{\hat\C}(\eta_1).
\end{equation}
In particular, we can choose $\eta=e^{-2\pi\tau}\bbS^1$, and write the above formula as the pairing:
\begin{equation}
    \frac{Z_k(\tau)}{Z_{\SLE}(\tau)} = e^{-\frac{\ccm}{2}\Lambda^*(\bbS^1,e^{-2\pi\tau}\bbS^1)} \left\langle \frac{Z_1(\tau_\eta)}{Z_{\SLE}(\tau_\eta)} \cB_1(\eta) \Big| \frac{Z_{k-1}(\tau_{\eta,e^{-2\pi\tau}\bbS^1})}{Z_{\SLE}(\tau_{\eta,e^{-2\pi\tau}\bbS^1})} \cB_2(\eta) \right\rangle_{\dd\nu_{\hat\C}(\eta)},
\end{equation}
where $\cB_1(\eta), \cB_2(\eta)$ are the boundary states defined in \eqref{eq:SLE-boundary-state}.

The two states in the above formula differ from the SLE boundary states $\cB_1, \cB_2$ only by factors of the conformal moduli. For any meromorphic vector field $\rv$ which is conformal in a neighborhood of the annulus and preserves $q\bbS^1$, the conformal modulus $\tau_{\eta,q\bbS^1}$ is invariant under the vector flow. Thus, for any smooth function on the modulus $m(\tau_{\eta,q\bbS^1})$, it commutes with the operators in \eqref{eq:boundary-state-contition-real}, and the boundary state condition \eqref{eq:boundary-state-condition} holds for $m(\tau_{\eta,q\bbS^1}) \cB_2(\eta)$. 

We can now do the same computation as in the previous subsection, and the only difference in the result appears in the one-point function. Recall that the one-point function comes from the pairing of the boundary state with the primary state, which in this case is given by:
\begin{equation}\label{eq:CLE-jth-1-pt-fct}
    \left\langle \frac{Z_j(\tau_\eta)}{Z_{\SLE}(\tau_\eta)} \cB_1(\eta) \Big| e^{-2\lambda c_\eta} \right\rangle_{\dd\nu_{\hat\C}(\eta)} = \CLE_{\bbD}[e^{-2\lambda c_{\eta_j}}] = C_1(\lambda)^j.
\end{equation}
Here $\eta_j$ with $j\ge1$ is the $j$-th outermost loop surrounding the origin in the $\CLE_\bbD$ configuration, and we use the domain Markov property of CLE. The one-point function is finite for $\mathrm{Re}(\ld) < -\frac{3\kappa}{64}-\frac{1}{\kappa}+\frac{1}{2} = \Delta_{(0,1/2)}$, but can be analytically extended to a meromorphic function for $\lambda\in\C$.
Its formula is rigorously proved in \cite{ssw-radii}:
\begin{equation}\label{eq:C_1}
    C_1(\lambda_P) = \frac{\cos(2\pi bP_{(1,1)})}{\cos(2\pi bP)}.
\end{equation}

The result is a similar formula as \eqref{eq:SLE-annulus-p-f}:
\begin{equation}\label{eq:CLE-k-annulus-p-f}
    \frac{Z_k(\tau)}{Z_{\SLE}(\tau)} = C\cdot \eta(2i\tau)^{\ccm}\tau^{\ccm/2} \int P\dd P\frac{C_1(\lambda_P)^k}{R_2(\lambda_P)} \frac{e^{-4P^2\pi\tau}}{\eta(2i\tau)}.
\end{equation}
Our derivation relies on the assumption $k\ge2$, which corresponds to the case that there exists a non-contractible loop in the annulus. The reason is that the Hilbert space we consider is the function space on a single SLE loop and we need to assume the existence of the loop to apply the spectral resolution computation. However, we can directly extend the formula \eqref{eq:CLE-k-annulus-p-f} to $k=1$. We can give the following interpretation: in this case, the second boundary state formally degenerates to the ``Dirac delta function'' that restricts $\eta$ to be $q\bbS^1$. Thus, the one-point function corresponding to this boundary state should be the moment of the conformal radius of the boundary:
\[ \CLE_{\bbD}[e^{-2\lambda\log\CR(\bbS^1;0)}] = 1 = C_1(\lambda)^0, \]
which is consistent with \eqref{eq:CLE-jth-1-pt-fct} for $j=0$.

Now we can give the $O(n,n')$ partition function. We write
\[ n'=2\cos(2\pi bP'). \]
The partition function is:
\begin{equation}
\begin{split}
    & \frac{Z(\tau;n')}{Z_{\SLE}(\tau)} = \sum_{k\ge1} \left(\frac{n'}{n}\right)^{k-1} \frac{Z_k(\tau)}{Z_{\SLE}(\tau)} \\
    =&\; C\cdot \eta(2i\tau)^{\ccm}\tau^{\ccm/2} \int P\dd P \sum_{k\ge1} \left(\frac{n'}{n}\right)^{k-1} \frac{C_1(\lambda_P)^k}{R_2(\lambda_P)} \frac{e^{-4P^2\pi\tau}}{\eta(2i\tau)} \\
    =&\; C\cdot \eta(2i\tau)^{\ccm}\tau^{\ccm/2} \int \frac{P\dd P}{R_2(\lambda_P)} \frac{C_1(\lambda_P)}{1-\frac{n'}{n}C_1(\lambda_P)} \frac{e^{-4P^2\pi\tau}}{\eta(2i\tau)} \\
    =&\; C\cdot \eta(2i\tau)^{\ccm}\tau^{\ccm/2} \int \dd P \sin(2\pi b^{-1}P) \frac{1}{\cos(2\pi bP)-\cos(2\pi bP')} \frac{e^{-4P^2\pi\tau}}{\eta(2i\tau)}. 
\end{split}
\end{equation}
The integrand has poles at $\pm P'+mb^{-1}$ for $m\in\bbZ$. Note that the momentum $\pm P$ corresponds to the same weight $\lambda_P=\lambda_{-P}$, we will just consider the poles 
\begin{equation}
    P_m=P'+mb^{-1}, \qquad m\in\bbZ.
\end{equation}
By the Residue Theorem, we can write the partition function as 
\begin{equation}
    \frac{Z(\tau;n')}{Z_{\SLE}(\tau)}
    = C\cdot \eta(2i\tau)^{\ccm-1}\tau^{\ccm/2} \sum_{m\in\bbZ} \frac{\sin(2\pi b^{-1}P_m)}{\sin(2\pi bP')} e^{-4P_m^2\pi\tau}.
\end{equation}

The SLE moduli distribution can be computed rigorously based on the variational methods and LQG coupling technique developed in \cite{BJSW-inprep}:
\begin{equation}
    Z_\SLE(\tau) = C\cdot \eta(2i\tau)^{1-\ccm} \tau^{-\ccm/2} e^{4P_{(1,1)}^2\pi\tau}.
\end{equation}
Thus, the $O(n,n')$ annulus partition function is:
\begin{equation}\label{eq:O(n,n')-partition-function}
    Z(\tau;n') = C\cdot \sum_{m\in\bbZ} \frac{\sin(2\pi b^{-1}P_m)}{\sin(2\pi bP')} e^{-4\lambda_{P_m}\pi\tau}.
\end{equation}
This matches with the formula in \cite{ars-annuli}, and the momentum set $\{P_m=P'+mb^{-1}:m\in\bbZ\}$ matches with the physical closed-channel spectrum.

A key point in the derivation is the determination of the spectrum. In our derivation, we do not a priori determine the spectrum. Instead, we formally choose an arbitrary continuum contour as an auxiliary spectral line, and use the Residue Theorem in the final step to make this continuum integral into a discrete summation over the genuine physical spectrum. The poles of the integrand come from the geometric-series summation of structure constants, and depend on the specific loop model. We will also see the same mechanism determining the spectrum for other loop models and for the disk two-point functions in the following sections.

\begin{remark}
It is implicit in the proof in \cite{ars-annuli} that the formula for the SLE one-point function $S_1(\lambda)$ can be determined by the moduli distribution $Z_\SLE(\tau)$, and vice versa. However, it is not easy to derive either of them. Our analytic bootstrap computation gives only the ratio $Z(\tau;n')/Z_{\SLE}(\tau)$, and combined with the formula $Z(\tau;n')$ in \cite{ars-annuli} can be viewed as an alternative physical derivation of $Z_\SLE(\tau)$.
\end{remark}

\begin{remark}
The SLE-loop CFT developed in Section~\ref{sec:CFT-of-SLE} is currently available only for simple loops, namely for $\CLE_\kappa$ with $\kappa\in(8/3,4]$, corresponding to dilute $O(n)$ models. In later sections, we use the same closed-channel formalism as a natural extension to the non-simple regime, namely to $\CLE_{\kappa'}$ with $\kappa'\in(4,8)$, corresponding to dense $O(n)$, FK and Potts models. We do not currently know how to justify this extension rigorously within the SLE-loop Hilbert-space framework. Nevertheless, the resulting formulae provide a consistency check, since they agree with the physical predictions.
\end{remark}

\subsection{FK/Potts models with boundary-touching loop reweightings}\label{subsec:JS-annulus}
We now consider the annulus loop model with boundary-touching loop reweightings as introduced in \cite{DJS2009}. We will consider the Potts models or FK random cluster models on the annulus, which correspond to non-simple $\CLE_{\kappa'}$ for $\kappa'\in (4,8)$. The parameters are as follows:
\begin{align*}
    & n = 2\cos(2\pi bP_{(1,1)}), && n_j = \frac{\sin(2\pi b(p_j+P_{(1,1)}))}{\sin(2\pi bp_j)}, &&& j=1,2, \\
    & l=2\cos(2\pi bP_l), && l_j = \frac{\sin(4\pi(y_j+\frac{b}{2}P_l))}{\sin(4\pi y_j)}, &&& j=1,2.
\end{align*}
In the lattice model, the loops come with 7 weights, depending on their topology and on touching the two boundaries. Contractible loops not touching either boundary have weight $n$, contractible loops touching only the $j$-th boundary have weight $n_j$, non-contractible loops not touching either boundary have weight $l$, non-contractible loops touching only the $j$-th boundary have weight $l_j$. In addition, we restrict to the case that there is no contractible loop touching both boundaries, or equivalently $n_{12}=0$. Note that in the FK models, each loop is an interface of a genuine cluster and a dual cluster, and thus the number of non-contractible loops is even. This parity constraint implies that there is at most one loop touching each boundary and no loop touching both boundaries.

The scaling limit of this loop model is conjectured to be described by $\mathrm{BCLE}$. The partition function consists of four parts, depending on whether there exists a non-contractible loop touching either boundary. For each boundary, if there is a non-contractible loop touching it, then the loop is described by the ``true loop'' in the BCLE language. Otherwise, we should consider the outer boundary enclosing all the boundary-touching loops, which corresponds to the ``false loop'' in the BCLE language. Thus, for each boundary component, we can associate a loop to it, and the configuration inside is the ordinary $O(n,l)$ model as described in the previous section, with the parity constraint. Both the ``true loop'' and the ``false loop'' are SLE type loops, and their corresponding disk 1-point function is computed in \cite{liu2024bulkonearmexponentclekappa}:
\begin{equation}
\begin{split}
    C_1^{\mathrm{true}}(\ld_P;n_j) = n_j^{-1} \mathbb{E}[\CR(0, D_\mathcal{L})^{-2\lambda_P} \mathbf{1}_{0 \in \mathrm{BCLE}_{\kappa',\rho_j}^\circlearrowright}] = C(n_j) \frac{\sin\left(2\pi b^{-1}P\left(1-\frac{\rho_j+2}{2}b^2\right)\right)}{\sin(2\pi b^{-1}P)} 
    = C(n_j) \frac{\sin(-4\pi p_jP)}{\sin(2\pi b^{-1}P)}, \\
    C_1^{\mathrm{false}}(\ld_P;n_j) = \mathbb{E}[\CR(0, D_\mathcal{L})^{-2\lambda_P} \mathbf{1}_{0 \notin \mathrm{BCLE}_{\kappa',\rho_j}^\circlearrowright}] = C(n_j)\frac{\sin\left(2\pi b^{-1}P\left(-1+\frac{\rho_j+4}{2}b^2\right)\right)}{\sin(2\pi b^{-1}P)} = C(n_j)\frac{\sin(4\pi(p_j+\frac{b}{2})P)}{\sin(2\pi b^{-1}P)}.
\end{split}
\end{equation}

The partition function is:
\begin{equation}
\begin{split}
    & \int \frac{P\dd P}{R_2(\ld_P)} \sum_{k\ge0} \left(\frac{l}{n}\right)^{2k} l_1l_2 C_1^{2k+1}(\ld_P) C_1^{\mathrm{true}}(\ld_P;n_1) C_1^{\mathrm{true}}(\ld_P;n_2) e^{-4\ld_P\pi\tau} \\
    +& \int \frac{P\dd P}{R_2(\ld_P)} \sum_{k\ge0} \left(\frac{l}{n}\right)^{2k+1} l_1 C_1^{2k+2}(\ld_P) C_1^{\mathrm{true}}(\ld_P;n_1) C_1^{\mathrm{false}}(\ld_P;n_2) e^{-4\ld_P\pi\tau} \\
    +& \int \frac{P\dd P}{R_2(\ld_P)} \sum_{k\ge0} \left(\frac{l}{n}\right)^{2k+1} l_2 C_1^{2k+2}(\ld_P) C_1^{\mathrm{false}}(\ld_P;n_1) C_1^{\mathrm{true}}(\ld_P;n_2) e^{-4\ld_P\pi\tau} \\
    +& \int \frac{P\dd P}{R_2(\ld_P)} \sum_{k\ge0} \left(\frac{l}{n}\right)^{2k}C_1^{2k+1}(\ld_P) C_1^{\mathrm{false}}(\ld_P;n_1) C_1^{\mathrm{false}}(\ld_P;n_2) e^{-4\ld_P\pi\tau},
\end{split}
\end{equation}
which is equal to a factor depending on the loop weights times
\begin{equation}
\begin{split}
    & \int \dd P \frac{\cos(2\pi bP)\cos(2\pi bP_l)}{\cos^2(2\pi bP)-\cos^2(2\pi bP_l)} \frac{\sin(-4\pi p_1P) \sin(-4\pi p_2P)}{\sin(2\pi b^{-1}P)} \frac{\sin(4\pi(y_1+\frac{b}{2}P_l))}{\sin(4\pi y_1)} \frac{\sin(4\pi(y_2+\frac{b}{2}P_l))}{\sin(4\pi y_2)} e^{-4\ld_P\pi\tau} \\
    &+ \int \dd P \frac{\cos^2(2\pi bP_l)}{\cos^2(2\pi bP)-\cos^2(2\pi bP_l)} \frac{\sin(-4\pi p_1P) \sin(4\pi(p_2+\frac{b}{2})P)}{\sin(2\pi b^{-1}P)} \frac{\sin(4\pi(y_1+\frac{b}{2}P_l))}{\sin(4\pi y_1)} e^{-4\ld_P\pi\tau} \\
    &+ \int \dd P \frac{\cos^2(2\pi bP_l)}{\cos^2(2\pi bP)-\cos^2(2\pi bP_l)} \frac{\sin(4\pi(p_1+\frac{b}{2})P) \sin(-4\pi p_2P)}{\sin(2\pi b^{-1}P)} \frac{\sin(4\pi(y_2+\frac{b}{2}P_l))}{\sin(4\pi y_2)} e^{-4\ld_P\pi\tau} \\
    &+ \int \dd P \frac{\cos(2\pi bP)\cos(2\pi bP_l)}{\cos^2(2\pi bP)-\cos^2(2\pi bP_l)} \frac{\sin(4\pi(p_1+\frac{b}{2})P) \sin(4\pi(p_2+\frac{b}{2})P)}{\sin(2\pi b^{-1}P)} e^{-4\ld_P\pi\tau}.
\end{split}
\end{equation}
The poles of the integrand consist of two parts: $P=P_l+kb^{-1}$ for $k\in\frac{1}{2}\bbZ$, and $P=P_{(r,0)}=\frac{rb}{2}$ for $r\in\bbZ_+$.
By some trigonometric function identities, the residue summation contributed by the first part is:
\begin{equation*}
    -\cos(2\pi bP_l)\sin(2\pi bP_l) \sum_{k\in\frac{1}{2}\bbZ} \frac{(-)^{2k}}{4\pi b} \frac{\sin\left(4\pi(y_1-p_1(P_l+kb^{-1}))\right) \sin\left(4\pi(y_2-p_2(P_l+kb^{-1}))\right)}{\sin(2\pi b^{-1}(P_l+kb^{-1})) \cdot \sin(4\pi y_1) \sin(4\pi y_2)} e^{-4\ld_{P_l+kb^{-1}}\pi\tau},
\end{equation*}
and the residue summation contributed by the second part is:
\begin{equation*}
    -\cos(2\pi bP_l)\sum_{r\in\bbZ\setminus\{0\}} \frac{b}{4\pi} \left[ \cos(4\pi p_{12}P_{(r,0)}) - (-)^r \frac{\sin(4\pi(y_1+p_1P_{(r,0)})) \sin(4\pi(y_2+p_2P_{(r,0)})) \sin(2\pi bP_l)}{\sin(2\pi b(P_{(r,0)}+P_l)) \sin(4\pi y_1) \sin(4\pi y_2)} \right] e^{-4\Delta_{(r,0)}\pi\tau},
\end{equation*}
where
\[ p_{12}=p_1+p_2+P_{(1,1)}, \]
that corresponds to the assumption $n_{12}=0$.
Adding these two parts together, we get:
\begin{equation}
\begin{split}
    \sum_{k\in\frac{1}{2}\bbZ} \frac{(-)^{2k}}{b} \frac{\sin\left(4\pi(y_1-p_1(P_l+kb^{-1}))\right) \sin\left(4\pi(y_2-p_2(P_l+kb^{-1}))\right) \sin(2\pi bP_l)}{\sin(2\pi b^{-1}(P_l+kb^{-1})) \cdot \sin(4\pi y_1) \sin(4\pi y_2)} e^{-4\ld_{P_l+kb^{-1}}\pi\tau} \\
    +\sum_{r\in\bbZ\setminus\{0\}} b \left[ \cos(4\pi p_{12}P_{(r,0)}) - (-)^r \frac{\sin(4\pi(y_1+p_1P_{(r,0)})) \sin(4\pi(y_2+p_2P_{(r,0)})) \sin(2\pi bP_l)}{\sin(2\pi b(P_{(r,0)}+P_l)) \sin(4\pi y_1) \sin(4\pi y_2)} \right] e^{-4\Delta_{(r,0)}\pi\tau}.
\end{split}
\end{equation}

\section{Bulk two-point functions in the disk}\label{sec:disk-two-point-functions}

\subsection{Disk two-point amplitudes}\label{subsec:disk-2-pt-amplitude}
In this section, we define the bulk two-point amplitudes and study the descendant Ward identities. We restrict to the dilute $O(n)$ model, and discuss in the later subsections the extension to other models.

\smallskip

\paragraph{Construction of the two-point disk amplitudes.}
Fix $z_1=0,z_2=z\in\bbC$, and a loop $\eta$ separating $0,z$ from $\infty$. Recall that $c_\eta$ and $\tilde c_\eta$ are logarithmic conformal radii of $\eta$ viewed from $0$ and $\infty$. We denote the law of the CLE inside $\eta$ by $\CLE_{D_\eta}$, which is equivalent to $\CLE_\bbD$ through the uniformizing map, and denote its configuration by $\Gamma$. For $j=1,2$, there are infinitely many loops in $\Gamma$ surrounding $z_j$ almost surely, and we denote by $\eta_j$ the outermost one surrounding only $z_j$ but not the other one and write $c_j:=\log\CR(\eta_j;z_j)$. Let $E$ be the topological event that there is no loop in $\Gamma$ surrounding both $0$ and $z$. For any weights $\lambda_1,\lambda_2$, we define the two-point disk amplitudes as:
\begin{equation}\label{eq:def-2-pt-amplitude}
    \cA(\eta;0,z;\ld_1,\ld_2) := \CLE_{D_\eta}\left[ \ind_E e^{-2\ld_1c_1} e^{-2\ld_2c_2} \right].
\end{equation}
In the discrete lattice $O(n)$ model, this can be viewed as reweighting the loops surrounding only $z_j$ by $\frac{n_j}{n}$ with the parameters related as follows:
\[ n_j = 2\cos(2\pi bP_j), \qquad \lambda_j = \lambda_{P_j} = P_j^2-P_{(1,1)}^2. \]

The disk two-point functions can be written as the pairing of the two-point disk amplitudes $\cA(\eta;0,z;\ld_1,\ld_2)$ with certain boundary states. In the previous section we have decomposed the boundary states into descendant states, and in the following we compute the pairing of $\cA(\eta;0,z;\ld_1,\ld_2)$ with descendant states.

\paragraph{Pairing with primary states.}
The pairing of $\cA(\eta;0,z;\ld_1,\ld_2)$ with primary states is given by the sphere 3-point functions. To see this, we first recall the sphere 3-point functions from the sphere CLE. Let $\Gamma$ be a configuration of $\CLE_{\hat\bbC}$, and denote by $\mathrm{Count}_\Gamma[\dd\eta]$ the counting measure of all loops in $\Gamma$.
As proved in \cite{ACSW24SLEloop}, if we first sample a probabilistic CLE configuration $\Gamma$ on $\hat\bbC$, and choose by counting measure a loop $\eta$ in $\Gamma$ that separates $0$ from $\infty$, then the law of $\eta$ coincides with the SLE measure on $\hat\bbC$ up to a normalizing constant:
\begin{equation}
    \int \CLE_{\hat\bbC}[\dd\Gamma] \Bigl( \mathrm{Count}_\Gamma[\dd\eta] \ind_{\eta\in\cJ_{0,\infty}} \Bigr) = \mathcal{D} \cdot \nu_{\hat\bbC}[\dd\eta], \qquad \mathcal{D}=\frac{1-b^{-2}}{\sqrt{\pi}\tan(\pi b^2)}
\end{equation}
For $z_1=0,z_2=z,z_3=\infty$, let $\eta_j$ be the outermost loop in $\Gamma$ surrounding only $z_j$ but not the other two points, and write $c_j=\log\CR(\eta_j;z_j)$. The conformal invariance of $\CLE_{\hat\bbC}$ constrains the 3-point to be
\begin{equation}\label{eq:CLE-3pt-fct}
    \CLE_{\hat\bbC}\left[\prod_{j=1}^3 e^{-2\lambda_jc_j}\right] = C_3(\lambda_1,\lambda_2,\lambda_3) |z|^{-2(\lambda_1+\lambda_2-\lambda_3)},
\end{equation}
and the structure constants are determined in \cite{ACSW24ImDOZZ}:
\begin{equation}
    C_3(\lambda_1,\lambda_2,\lambda_3) = C\cdot \prod_{j=1}^3 \frac{\sin(2\pi bP_j)}{\cos(2\pi bP_j)} \frac{\prod_{j=1}^3 \left(P_j\prod_{\pm}\Gamma_b(\pm2P_j)\right)}{\prod_{\pm,\pm,\pm}\Gamma_b\left(\frac{Q}{2}\pm P_1\pm P_2\pm P_3\right)}.
\end{equation}
We also introduce the following convention
\begin{equation}
    \hat C_{P_1,P_2}^P = \frac{16P \sin(\pi b^2)\sin(\pi b^{-2})\Gamma_b(2P)\Gamma_b(-2P)}{\prod_{\epsilon_1, \epsilon_2, \epsilon_3 = \pm} \Gamma_b\left(\frac{b + b^{-1}}{2} +\epsilon_1P_1+\epsilon_2P_2 + \epsilon_3 P\right)},
\end{equation}
and thus the structure constants are:
\begin{equation}
    C_3(\lambda_{P_1},\lambda_{P_2},\lambda_{P}) = \mathcal N_0(P_1,P_2) \cdot \frac{\sin(2\pi bP)}{\cos(2\pi bP)} \hat{C}^P_{P_1,P_2},
\end{equation}
with constant
\[ \mathcal N_0(P_1,P_2) = -\frac{2b^{b^{-2}-b^2}\Gamma_b(Q)^2}{(b-b^{-1})\tan^3(\pi b^2) \Gamma(1-b^2)\Gamma(1-b^{-2})} \prod_{j=1}^{2} \left[ P_j\tan(2\pi bP_j) \Gamma_b(2P_j)\Gamma_b(-2P_j) \right]. \]

We can rewrite the 3-point function \eqref{eq:CLE-3pt-fct} in the following way. Note that the outermost loop $\eta_3$ surrounding $z_3=\infty$ is the unique loop in $\Gamma$ such that it separates $0$ from $\infty$ and that there is no other loop inside it surrounding both $0$ and $z$. The first condition restricts the SLE measure on $\hat\bbC$ to $\nu_{\hat\bbC}$, and the second condition is the same as the topological event $E$ in \eqref{eq:def-2-pt-amplitude}. In addition, the domain Markov property guarantees that conditioned on the loop $\eta_3$, the loop configuration inside $\eta_3$ is an independent $\CLE_{D_\eta}$. Thus, we can rewrite \eqref{eq:CLE-3pt-fct} as:
\begin{equation}
\begin{split}
    & \CLE_{\hat\bbC}\left[\prod_{j=1}^3 e^{-2\lambda_jc_j}\right]
    = \int \CLE_{\hat\bbC}[\dd\Gamma] \mathrm{Count}_\Gamma[\dd\eta_3] \ind_{\eta_3\in\cJ_{0,\infty}} e^{-2\lambda_3 c_3}\CLE_{D_\eta}\left[ \ind_E e^{-2\ld_1c_1} e^{-2\ld_2c_2} \right] \\
    ={}& \mathcal{D}\cdot \int \dd\nu_{\hat\bbC}(\eta_3) e^{-2\lambda_3c_3}\cA(\eta_3;0,z;\ld_1,\ld_2)
    = \mathcal{D}\cdot \left\langle e^{-2\lambda_3\tilde c_\eta} \Big| \cA(\eta;0,z;\ld_1,\ld_2) \right\rangle_{\dd\nu_{\hat\C}(\eta)}.
\end{split}
\end{equation}
Compared with \eqref{eq:CLE-3pt-fct}, we derive the pairing of the 2-point amplitudes with primary states:
\begin{equation}\label{eq:pairing-2-pt-amplitude-primary}
    \left\langle e^{-2\lambda\tilde c_\eta} \Big| \cA(\eta;0,z;\ld_1,\ld_2) \right\rangle_{\dd\nu_{\hat\C}(\eta)} = \mathcal{D}^{-1} \cdot C_3(\lambda_1,\lambda_2,\lambda) |z|^{-2(\lambda_1+\lambda_2-\lambda)}.
\end{equation}

\paragraph{Pairing with descendant states.}
The Ward identities for $\cA(\eta;0,z;\ld_1,\ld_2)$ follow directly from the conformal invariance of CLE. For any vector field $\rv=v(z)\partial_z\in\C[z]$, which is holomorphic on $\bbC$ with pole at $\infty$, consider the vector flow $\phi_t(z)=e^{t\rv}(z)=z+tv(z)+o(t)$. For $t\in\bbC$ small enough, $\phi_t(z)$ is an injective conformal map in a neighborhood of $\intt(\eta)$. By the conformal invariance, the $\CLE_\eta$ and $\CLE_{\phi_t(\eta)}$ are equivalent through the conformal map $\phi_t$, and the local operators $e^{-2\lambda_jc_j}$ transform covariantly. This gives:
\begin{equation}\label{eq:conf-cov-2-pt-amplitude}
    \cA(\phi_t(\eta);\phi_t(z_1),\phi_t(z_2);\ld_1,\ld_2) \cdot |\phi_t'(z_1)|^{2\lambda_1} |\phi_t'(z_2)|^{2\lambda_2} = \cA(\eta;z_1,z_2;\ld_1,\ld_2).
\end{equation}
Infinitesimally, this gives the Ward identity:
\begin{equation}\label{eq:Ward-2-pt-amplitude}
\begin{split}
    \left( \bL^{(3)}_{\rv,\ext} + \sum_{j=1,2} \left(v(z_j)\pt_{z_j} + \lambda_j v'(z_j)\right) \right) \cA(\eta;z_1,z_2;\ld_1,\ld_2) = 0, \\
    \left( \bbL^{(3)}_{\rv,\ext} + \sum_{j=1,2} \left(\ol{v(z_j)}\bar\pt_{z_j} + \lambda_j\ol{v'(z_j)}\right) \right) \cA(\eta;z_1,z_2;\ld_1,\ld_2) = 0.
\end{split}
\end{equation}
Here the Virasoro operators $\bL^{(3)}_{\rv,\ext}$ and $\bbL^{(3)}_{\rv,\ext}$ act on $\eta$, and we view the component of $\hat\bbC\setminus\eta$ containing $0$ as its exterior (see Subsection \ref{subsec:multi-loop-measure-Ward}).

Define
\begin{equation}
    \bD_n = z^{n+1}\pt_z + (n+1)\ld_2z^n, \qquad \bar\bD_n = \bar z^{n+1}\pt_{\bar z} + (n+1)\bar\ld_2\bar z^n.
\end{equation}
For $n\ge 1$, we have $v_n(0)=v_n'(0)=0$. Thus we have
\begin{equation}
    \left( \bL_\rv + \bD_n \right) \cA(\eta;0,z;\ld_1,\ld_2) = 0, \qquad \left( \bbL_\rv + \bar\bD_n \right) \cA(\eta;0,z;\ld_1,\ld_2) = 0.
\end{equation}

For any function $F(\eta)$ and $n\ge1$, the Virasoro adjoint relation and the Ward identity gives:
\begin{equation}
\begin{split}
    \left\langle \bL^{(3)}_{-n,\intt}F(\eta) \big|\cA(\eta;0,z;\ld_1,\ld_2) \right\rangle_{\dd\nu_{\hat\C}(\eta)} = \bD_n\left\langle F(\eta) \big|\cA(\eta;0,z;\ld_1,\ld_2) \right\rangle_{\dd\nu_{\hat\C}(\eta)}, \\
    \left\langle \bbL^{(3)}_{-n,\intt}F(\eta) \big|\cA(\eta;0,z;\ld_1,\ld_2) \right\rangle_{\dd\nu_{\hat\C}(\eta)} = \bar\bD_n\left\langle F(\eta) \big|\cA(\eta;0,z;\ld_1,\ld_2) \right\rangle_{\dd\nu_{\hat\C}(\eta)}.
\end{split}
\end{equation}
We now choose $F(\eta)$ to be the primary states $e^{-2\lambda_3\tilde c_\eta}$ and apply the above formula iteratively. Note the 3-point function \eqref{eq:pairing-2-pt-amplitude-primary} is a simple power function in $|z|$, whose derivatives can be computed easily. For any Young diagram $\nu = (\nu_i)_{i\ge1}$, define
\begin{equation}\label{eq:Ward_coefficient}
    v(\ld_1,\ld_2,\ld_3,\nu) := \prod_{j\ge 1} (\nu_j\ld_2 - \ld_1 + \ld_3 + \sum_{u<j}\nu_u).
\end{equation}
We can now compute the pairing of $\cA(\eta;0,z;\ld_1,\ld_2)$ with descendant states inductively:
\begin{equation}
\begin{split}
    & \left\langle \bL_{-\nu,\rm\intt}^{(3)} \bbL_{-\tilde\nu,\rm\intt}^{(3)} e^{-2\ld \tilde c_\eta} \big| \cA(\eta;0,z;\ld_1,\ld_2) \right\rangle_{\dd\nu_{\hat\C}(\eta)}\\
    ={}& \mathcal{D}^{-1} \cdot v(\ld_1,\ld_2,\ld,\nu) v(\ld_1,\ld_2,\ld,\tilde\nu) z^{|\nu|} \bar z^{|\tilde\nu|} |z|^{-2(\ld_1+\ld_2-\ld)} C_3^\CLE(\ld_1,\ld_2,\ld).
\end{split}
\end{equation}

\subsection{Closed-channel formula for disk two-point functions} \label{subsec:disk-2-pt-fct}
We now consider the two-point functions introduced in Section~\ref{subsec:intro-result}. We first consider the contribution by configurations with fixed number of common-surrounding loops $N_{\mathrm{s}}=k\ge0$:
\begin{equation}\label{eq:disk-observable-k}
    \cG^{(k)}_\bbD(P_1,P_2;z) := \bbE_{\CLE_\bbD}\!\left[ e^{-2\lambda_{P_1}c_1}e^{-2\lambda_{P_2}c_2} \ind_{N_{\mathrm{s}}=k} \right]
\end{equation}

We start by decomposing the correlation function as the pairing of states of a loop, using the domain Markov property of CLE.
Let $\eta$ be the innermost loop surrounding both marked points, and set $\eta$ to be $\bbS^1$ if there is no such loop. For the same reason discussed in Section~\ref{subsection:annulus-O(n)-p-f}, we will first demonstrate the computation for $k\ge1$ and then extend the resulting formula to $k=0$.

On the event $\{N_{\mathrm{s}}=k\}$, the law of $\eta$ is the same as following: choose the $k$-th outermost loop in $\Gamma$ surrounding $0$, and then restrict to the event that $z$ is inside the loop. It is explained in Section~\ref{subsection:annulus-O(n)-p-f} that the law of this $k$-th outermost CLE loop is related to the SLE loop measure by the boundary states
\[ \eta \sim \frac{Z_k(\tau_\eta)}{Z_{\SLE}(\tau_\eta)} e^{\frac{\ccm}{2}\Lambda^*(\eta,\bbS^1)} \dd\nu_{\hat\C}(\eta). \]
By the domain Markov property of CLE, the loop configuration inside $\eta$ is a CLE in the subdomain, independent of the configuration outside $\eta$, and we should restrict to the event that there is no further loop inside $\eta$ surrounding both marked points. Thus, we can decompose the correlation as
\begin{equation*}
\begin{split}
    &\cG^{(k)}_\bbD(P_1,P_2;z) \\
    ={}& \left\langle \frac{Z_k(\tau_\eta)}{Z_{\SLE}(\tau_\eta)} e^{\frac{\ccm}{2}\Lambda^*(\eta,\bbS^1)} \Big| \cA(\eta;0,z;\ld_1,\ld_2) \right\rangle_{\dd\nu_{\hat\C}(\eta)}.
\end{split}
\end{equation*}
The spectral resolution of the boundary state in Section \ref{sec:annulus} gives:
\begin{equation}
    \frac{Z_k(\tau_\eta)}{Z_{\SLE}(\tau_\eta)} e^{\frac{\ccm}{2}\Lambda^*(\eta,\bbS^1)} = \int_\bbR \frac{\dd p}{\sqrt{\pi}} \sum_{|\nu|=|\tilde\nu|} B_{\lambda_0+ip}^{-1}(\tilde\nu,\nu) \frac{C_1(\lambda_0+ip)^k}{R_2(\lambda_0+ip)} \bL^{(3)}_{-\nu,\intt}\bbL^{(3)}_{-\tilde\nu,\intt}(e^{-2(\lambda_0+ip)\tilde c_\eta}).
\end{equation}
Combining the pairing of the disk 2-point amplitudes with primaries and descendants computed in Section~\ref{subsec:disk-2-pt-fct}, we have
\begin{equation}\label{eq:fix-k-2-pt-formula}
\begin{split}
    &\cG^{(k)}_\bbD(P_1,P_2;z)
    = \mathcal{D}^{-1} \cdot \int_\bbR \frac{\dd p}{\sqrt{\pi}} \frac{C_1(\lambda_0+ip)^k}{R_2(\lambda_0+ip)} C_3(\lambda_{P_1},\lambda_{P_2},\lambda_0+ip) \cF^{\ccm}_{\lambda_0+ip}(\lambda_{P_1},\lambda_{P_2},\lambda_{P_2},\lambda_{P_1}; |z|^2) \\
    ={}& \mathcal N_1(P_1,P_2) \left(\frac{n}{2}\right)^k
    \int_{\mathcal C}\!\dd\lambda\, \frac{\sin\bigl(2\pi b^{-1}P_\lambda\bigr)\sin\bigl(2\pi bP_\lambda\bigr)}{P_\lambda\cos^{k+1}\bigl(2\pi bP_\lambda\bigr)} \hat{C}^{P_\lambda}_{P_1,P_2} \cF^{\ccm}_{\lambda}(\lambda_{P_1},\lambda_{P_2},\lambda_{P_2},\lambda_{P_1};|z|^2),
\end{split}
\end{equation}
where 
\[ P_\lambda = \sqrt{\lambda+\frac{(b-b^{-1})^2}{4}} \]
is the momentum for the conformal weight $\lambda$ (the integrand is independent of the choice of the square root),
$\cF$ is the sphere 4-point conformal block:
\begin{equation}\label{eq:block-definition}
    \cF^{\ccm}_{\ld}(\ld_1,\ld_2,\ld_3,\ld_4; z) = z^{\ld-\ld_1-\ld_2} \sum_{n\ge0} \sum_{|\nu|=|\tilde\nu|=n} v(\ld_1,\ld_2,\ld,\tilde\nu) v(\ld_4,\ld_3,\ld,\nu) B_{\ld}^{-1}(\nu,\tilde\nu) z^{n}.
\end{equation}
Recall that $v$ is the algebraic coefficient from the Ward identity defined in \eqref{eq:Ward_coefficient}.

The constant $\mathcal N_1(P_1,P_2)$ is independent of $z$ whose explicit expression is:
\[ \mathcal N_1(P_1,P_2) = \frac{ib^{b^{-2}-b^2}\Gamma_b(Q)^2}{(1-b^{-2}) \sin(\pi b^{-2}) \tan^2(\pi b^2) \Gamma(1-b^2)\Gamma(1-b^{-2})} \prod_{j=1}^{2} \left[ P_j\tan(2\pi bP_j) \Gamma_b(2P_j)\Gamma_b(-2P_j) \right]. \]

The straightforward choice of the contour line is the vertical line $\mathcal{C}=\lambda_0+i\bbR$ for $\lambda_0<-P_{(1,1)}^2$.
The convergence of the integral along this vertical line in \eqref{eq:fix-k-2-pt-formula} is valid for $k\ge1$. In the formal extension to $k=0$, the integral fails to converge along the vertical spectral line in the dense regime $\kappa > 4$, $b < 1$. 
The asymptotic behavior of the Barnes structure constant is
\[ \left| \hat{C}^P_{P_1,P_2} \right| \sim C \cdot |P^\nu| e^{-\pi(b+b^{-1})\mathrm{Im}(P)} |16^{-P^2}|, \qquad |P|\to\infty \]
for some explicit constant $C$ and $\nu=4(\lambda_{P_1}+\lambda_{P_2})-\frac{\ccm+1}{4}$.
Let 
\[ q=q(|z|^2)=\exp\!\left[-\pi\frac{K(1-|z|^2)}{K(|z|^2)}\right]\in(0,1) \]
where $K$ is the complete elliptic integral of the first kind. The elliptic representation of the conformal block is \cite{Zamolodchikov1987}
\[ \cF^{\ccm}_{\lambda_P}(\lambda_{P_1},\lambda_{P_2},\lambda_{P_2},\lambda_{P_1};|z|^2) = \mathcal{E}(|z|^2) (16q)^{P^2} H_P(q),\]
with $\mathcal{E}(|z|^2)$ independent of $P$ and $H_P(q)=1+O(P^{-2})$.
Thus, the asymptotic behavior of the integrand in \eqref{eq:fix-k-2-pt-formula} is
\[ \sim C\cdot |P|^{\nu-1} q^{\mathrm{Re}(P^2)} e^{-\pi((2k+1)b-b^{-1})\mathrm{Im}(P)}. \]
On the vertical line $\lambda_P=P^2-P_{(1,1)}^2\in\lambda_0+i\bbR$, the term $q^{\mathrm{Re}(P^2)}=q^{\lambda_0}$ is a constant, and the integral converges if $(2k+1)b-b^{-1}>0$. This fails when $k=0$ and $b<1$ in the dense case.

However, we can deform the contour to get faster decay of the integrand. One natural choice is 
\begin{equation}
    \mathcal C_\delta: \lambda(u) = (u+i\delta)^2 - P_{(1,1)}^2, \qquad u\in\bbR,
\end{equation}
or equivalently, $P \in \bbR+i\delta$, for $\delta>0$. This contour line bent towards right on the complex $\lambda$-plane, and the shift $\delta>0$ is chosen to avoid singularities. A figure of it is shown in Figure~\ref{fig:parabolic-contour}.
Along this contour, the integrand has Gaussian decay as $u\to\infty$
\[ \sim C\cdot |u+i\delta|^{\nu-1} q^{u^2-\delta^2}, \]
and thus the integral \eqref{eq:fix-k-2-pt-formula} converges.

\begin{table}[htbp]
    \centering
    \begin{tabular}{cccc}
        \hline
        Regime & $k$ & Vertical contour $\lambda_0+i\mathbb R$
                     & Parabolic contour $\mathcal C_\delta$ \\
        \hline
        $8/3<\kappa<4$ & $k\geq 0$ & Yes & Yes \\
        $4<\kappa<8$   & $k\geq 1$ & Yes & Yes \\
        $4<\kappa<8$   & $k=0$     & Not in general & Yes \\
        \hline
    \end{tabular}
    \caption{Convergence of the fixed-$k$ integral
    in \eqref{eq:fix-k-2-pt-formula}.
    The case $\kappa=4$ is discussed separately.}
    \label{tab:contour-convergence}
\end{table}

Summing over $k$ with assigned weight $n_{\mathrm{s}}$ for each common-surrounding loop, we have
\begin{equation}
\begin{split}
    & \cG_\bbD(P_1,P_2,P_{\mathrm{s}};z) = \sum_{k\ge0} \left(\frac{n_{\mathrm{s}}}{n}\right)^k \cG^{(k)}_\bbD(P_1,P_2;z) \\
    ={}& \mathcal{D}^{-1}\cdot \sum_{k\ge0} \left(\frac{n_{\mathrm{s}}}{n}\right)^{k} \frac{1}{i\sqrt{\pi}}
    \int_{\mathcal C_\delta} \dd\lambda \frac{C_1(\lambda)^k}{R_2(\lambda)} C_3(\lambda_{P_1},\lambda_{P_2},\lambda) \cF^{\ccm}_{\lambda}(\lambda_{P_1},\lambda_{P_2},\lambda_{P_2},\lambda_{P_1}; |z|^2) \\
    ={}& \mathcal{D}^{-1}\cdot \frac{1}{i\sqrt{\pi}} \int_{\mathcal C_\delta} \dd\lambda \frac{C_3(\lambda_{P_1},\lambda_{P_2},\lambda)}{\left(1-\frac{n_{\mathrm{s}}}{n} C_1(\lambda)\right) \cdot R_2(\lambda)} \cF^{\ccm}_{\lambda}(\lambda_{P_1},\lambda_{P_2},\lambda_{P_2},\lambda_{P_1}; |z|^2) \\
    ={}& 2 \mathcal N_1(P_1,P_2) \cdot \int_{\bbR+i\delta} \dd P \frac{\sin(2\pi b^{-1}P) \sin(2\pi bP)}{\cos(2\pi bP)-\cos(2\pi bP_{\mathrm{s}})} \hat{C}^P_{P_1,P_2} \cF^{\ccm}_{\ld_P}(\lambda_{P_1},\lambda_{P_2},\lambda_{P_2},\lambda_{P_1}; |z|^2),
\end{split}
\end{equation}

As explained before, to transform the integral to a summation via a formal residue evaluation, we need to investigate the poles and residues of the integrand. The poles can be sorted into three parts. The first part is $P_m=P_{\mathrm{s}}+mb^{-1}$ for $m\in\bbZ$, and the residues are 
\[ -\frac{1}{2\pi b} \cdot \sin\left(2\pi b^{-1}P_m\right) \hat{C}^{P_m}_{P_1,P_2} \cF^{\ccm}_{\ld_{P_m}}(\lambda_{P_1},\lambda_{P_2},\lambda_{P_2},\lambda_{P_1}; |z|^2).  \]
The second part comes from the 3-point function $\hat{C}^P_{P_1,P_2}$ which has simple poles at $P_{(-m,n)}$ for $m,n\in\bbZ_{\ge0}$. The poles $P_{(m,0)}$ and $P_{(0,n)}$ are zeros of sine functions in the numerator, and thus do not contribute residues. The third part comes from the conformal block which has poles at $P_{(m,n)}$ for $m,n\in\bbZ_+$. A remarkable phenomenon is that for each $m,n\in\bbZ_+$, the residue at $P_{(-m,n)}$ cancels exactly with the residue at $P_{(m,n)}$. This was pointed out in \cite{Zamolodchikov2004HEM,SantachiaraViti2014LogCoulombGas,Ribault2019InterpolateMinimalModels}, and we record the computation in Appendix \ref{appendix:residue-cancellation}.

Applying the Residue Theorem, we get the formula for the disk two-point function:
\begin{equation}
\begin{split}
    \cG_\bbD(P_1,P_2,P_{\mathrm{s}};z) 
    = \mathcal N_2(P_1,P_2) \;\cdot & \sum_{m\in\bbZ} \sin\left(2\pi b^{-1}P_m\right) \hat{C}^{P_m}_{P_1,P_2} \cF^{\ccm}_{\ld_{P_m}}(\lambda_{P_1},\lambda_{P_2},\lambda_{P_2},\lambda_{P_1}; |z|^2),
\end{split}
\end{equation}
where the normalization constant is
\[ \mathcal N_2(P_1,P_2) = -\frac{2b^{b^{-2}-b^2-1}\Gamma_b(Q)^2}{(1-b^{-2}) \sin(\pi b^{-2}) \tan^2(\pi b^2) \Gamma(1-b^2)\Gamma(1-b^{-2})} \prod_{j=1}^{2} \left[ P_j\tan(2\pi bP_j) \Gamma_b(2P_j)\Gamma_b(-2P_j) \right]
\]

At $\kappa=4$ ($b=1$), the factor $\sin(\pi b^2)\sin(\pi b^{-2})$ in $\hat C$ has a double zero. For generic external momenta $P_1,P_2$, this zero cancels the double pole of each of $\mathcal N_1$ and $\mathcal N_2$. We therefore take the limit $b\to1$ in the products $\mathcal N_1\hat C$ and $\mathcal N_2\hat C$. In this interpretation, the integral \eqref{eq:fix-k-2-pt-formula} converges along the parabolic contour line $\mathcal{C}_\delta$.

\subsection{Two-point connectivity probabilities}\label{subsec:2-pt-connectivity}
In this section, we consider the disk two-point connectivity which was considered in \cite{DJNRS26}. Although in the previous section we focus on the dilute $O(n)$ model whose scaling limit is simple $\CLE_\kappa$ with $\kappa\in(8/3,4]$, we now analytically extend the formulae to dense $O(n)$ loop models and FK-$Q$ random cluster models which correspond to the non-simple $\CLE_{\kappa'}$ with
\begin{equation}
    \kappa' = \frac{4}{b^2} \in(4,8), \qquad Q=4\cos^2(\pi b^2).
\end{equation}
As derived in \cite{ssw-radii,ACSW24ImDOZZ,ACSW24SLEloop}, the structure constants for simple $\CLE_\kappa$ and non-simple $\CLE_{\kappa'}$ share a common analytic expression.

The connectivity correlation functions are defined in the continuum limit using the Miller-Schoug measure constructed in \cite{miller2023existence}, which is a $d$-dimensional measure on each CLE cluster with $d=2-2\Delta_{(0,\frac{1}{2})}$, and is conjectured to be the scaling limit of the volume measure on each cluster in lattice models. Let $\eta$ be a Jordan loop with interior $D_\eta$ and $z_1,z_2\in D_\eta$. The two-point connectivity amplitude is defined via the following disintegration: for any disjoint compact sets $U_1,U_2$ in $D_\eta$,
\[ \CLE_{D_\eta}\left[\iint_{U_1\times U_2} \mathcal{M}^{\mathrm{cluster}}(\dd z_1) \mathcal{M}^{\mathrm{cluster}}(\dd z_2)\right] = \iint_{U_1\times U_2} \cA^{\mathrm{conn}}(\eta;z_1,z_2)\dd z_1\dd z_2, \]
where $\mathcal{M}^{\mathrm{cluster}}$ is the Miller-Schoug measure for the outermost CLE cluster. In other words, $\cA^{\mathrm{conn}}(\eta;z_1,z_2)$ can be viewed as the scaling limit of the probability that both $z_1,z_2$ are in the unique cluster connected to the boundary $\eta$.

The conformal covariance of the Miller--Schoug measure implies that the connectivity amplitude $\cA^\mathrm{conn}(\eta;z_1,z_2)$ satisfies the same conformal covariance \eqref{eq:conf-cov-2-pt-amplitude} with $\lambda_1=\lambda_2=1-\frac{d}{2}=\Delta_{(0,\frac{1}{2})}$, and thus the Ward identity holds. The pairing of $\cA^\mathrm{conn}(\eta;z_1,z_2)$ with primary states can be rigorously derived by an easy generalization of the proof in \cite{ACSW24ImDOZZ}:
\begin{equation}
    \left\langle e^{-2\ld_P \tilde c_\eta} \mid \cA^\mathrm{conn}(\eta;0,z) \right\rangle_{\dd\nu_{\hat\C}(\eta)} = \mathcal N \cdot |z|^{-2(2\Delta_{(0,\frac{1}{2})}-\lambda_P)} \frac{\sin(2\pi bP)}{\cos(2\pi bP)} \hat{C}^P_{P_{(0,\frac{1}{2})},P_{(0,\frac{1}{2})}}.
\end{equation}
Here $\mathcal N$ is an overall constant depending on the normalization defining the Miller--Schoug measure.

We can extend the previous results to the disk two-point connectivity for $O(n)$ model:
\begin{equation}\label{eq:disk-2-pt-conn-formula}
\begin{split}
    &\; \cG_{\bbD, O(n)}^\mathrm{conn}(z) = \sum_{k\ge 0} \left\langle \frac{Z_k(\tau_\eta)}{Z_{\SLE}(\tau_\eta)} e^{\frac{\ccm}{2}\Lambda^*(\eta,\bbS^1)} \Big| \cA^\mathrm{conn}(\eta;0,z) \right\rangle_{\dd\nu_{\hat\C}(\eta)} \\
    =&\; \mathcal N' \cdot \sum_{k=0}^\infty \sin\left(2\pi b^{-1}P_{(1,2k+1)}\right) \hat{C}^{P_{(1,2k+1)}}_{P_{(0,\frac{1}{2})},P_{(0,\frac{1}{2})}} \cF^{\ccm}_{\Delta_{(1,2k+1)}}(\Delta_{(0,\frac{1}{2})},\Delta_{(0,\frac{1}{2})},\Delta_{(0,\frac{1}{2})},\Delta_{(0,\frac{1}{2})};|z|^2).
\end{split}
\end{equation}
Note that we have $\hat C^{(1,2k+1)}_{(0,\frac{1}{2}),(0,\frac{1}{2})}=0 $ for $k\le-1$ since there is factor $\Gamma_b((k+1)b^{-1})$ in the denominator, so only the $k\ge0$ part contributes to the residues summation. This gives the two-point connectivity for both dilute and dense $O(n)$ loop models.

In the FK-$Q$ random cluster models, the CLE clusters correspond to both open and closed clusters, which appear alternately. If we consider the free (resp. wired) boundary condition, then the scaling limit of FK clusters is the even-numbered (resp. odd-numbered) layer of $\CLE_{\kappa'}$ clusters. This corresponds to the odd $k$'s (resp. even $k$'s) in the summation in \eqref{eq:disk-2-pt-conn-formula} and we have
\begin{equation}
\begin{split}
    \cG_{\bbD,\mathrm{FK}}^{\mathrm{free(wired)}}(z) = \frac{1}{2}\mathcal N' \cdot \sum_{k=1}^\infty \epsilon^{(k-1)} \sin\left(2\pi b^{-1}P_{(1,k)}\right) \hat{C}^{P_{(1,k)}}_{P_{(0,\frac{1}{2})},P_{(0,\frac{1}{2})}} \cF^{\ccm}_{\Delta_{(1,k)}}(\Delta_{(0,\frac{1}{2})},\Delta_{(0,\frac{1}{2})},\Delta_{(0,\frac{1}{2})},\Delta_{(0,\frac{1}{2})};|z|^2),
\end{split}
\end{equation}
where $\epsilon=-1$ for free boundary conditions and $\epsilon=+1$ for wired boundary conditions.
The above formula matches with \cite[Formula (7)]{DJNS26}.

\section{Conclusion}\label{sec:conclusion}

\subsection{Summary}
We use the SLE-loop CFT to derive bulk two-point functions of critical loop models in the disk. 
The resulting formulae can be written as an integral over a certain spectral line involving Virasoro conformal blocks and, in some cases, can also be expressed as a summation over the discrete spectrum. The discrete summation formulae agree with the corresponding conformal bootstrap predictions for the $O(n)$ and the FK models \cite{DJNRS26,DJNS26}.
Reproducing known annulus partition functions provides a consistency check of the method.

We prove a local Ward identity for the Brownian-loop weight, which encodes the interaction term in a formal path integral for the multiple-SLE loop measure defined in Section~\ref{subsec:multi-loop-measure-Ward}. Joint laws of selected CLE loops can often be expressed as multiple-SLE loop measures weighted by functions of the conformal moduli of the domain bounded by the loops. Such representations can be established through the Liouville quantum gravity welding method \cite{ACSW24ImDOZZ,ars-annuli,BJSW-inprep}, and these moduli factors are encoded in the corresponding structure constants. The simultaneous local conformal deformations in the Ward identity preserve these moduli, so the corresponding Virasoro action commutes with multiplication by moduli factors. The Ward identity therefore remains valid in the presence of these factors.

The conformal bootstrap approach in \cite{Nivesvivat:2023kfp,DJNRS26} takes as input a bulk-channel spectrum obtained from Coulomb-gas and transfer-matrix calculations \cite{diFrancesco:1987qf,DJS2009,Jacobsen:2022nxs}. Our method does not assume this discrete spectrum in advance. The spectral resolution of Baverez--Jego involves a contour integral arising from the Fourier transform of the zero mode associated with the loop scale. For a general SLE observable, the resulting integral need not reduce to a discrete sum. In the models studied here, the explicit structure constants and the sum over the number of certain separating loops lead to a formal evaluation by residues. In the contour integral expression for disk two-point functions, the poles from the three-point structure constants and the Kac poles of the conformal blocks cancel pairwise by Zamolodchikov's recursion relation. In these examples, the remaining poles constituting the discrete spectrum are determined by the prescribed loop weights in the chosen expansion channel.

\subsection{Outlook}
Several steps in our derivation are formal. The spectral resolution in \cite{baverez2024cftsleloopmeasures} is established for ``sufficiently regular'' functions described in Section~\ref{sec:CFT-of-SLE}, and its rigorous extension to our boundary states and two-point amplitudes remains to be established. 
First, for the configurations with no separating loop in the specified channel, one needs to have a distributional interpretation for the boundary state, which formally constrains the loop to be the circle.
Second, since the Shapovalov form is not positive definite, one cannot directly use the usual Hilbert-space orthogonality argument to prove the convergence of the resulting formulae.
Estimates on the structure constants and conformal blocks are needed to justify the convergence of sums and integrals, and evaluating the integral by residues.

Another issue is the extension from the simple regime to the non-simple regime. The SLE-loop CFT used here is valid only for simple loops in $\CLE_\kappa$ with $\kappa\in(8/3,4]$, corresponding to the dilute $O(n)$ model. For dense $O(n)$ and FK models, we formally extend the resulting formulae to $\kappa'=4/b^2\in(4,8)$. 
The relevant CLE structure constants admit the same analytic expressions in the simple and non-simple regimes.
This motivates the formal continuation, but does not by itself establish a spectral resolution for non-simple SLE loop measures.

Two future goals are therefore to rigorously establish the spectral resolution for the CLE amplitudes and boundary states considered here, and to develop the corresponding CFT framework for non-simple loops. It would also be natural to consider more general boundary interactions (for example, the boundary-touching loop reweightings introduced in Section~\ref{subsec:JS-annulus}), and observables with more marked points.

For the FK/Potts annulus with boundary-touching loop weights introduced in Section~\ref{subsec:JS-annulus}, the present calculation excludes configurations containing contractible loops touching both boundaries. Such configurations do not contain a non-contractible loop, whereas our derivation relies on cutting along such a loop to apply the spectral resolution. 
It is natural to consider the extension to include these configurations.

Our approach is also useful for other critical loop models whose interfaces admit an SLE or CLE description. Applying it to such models would require identifying the law of relevant multi-loop measures in terms of multiple SLE measures and the associated model-dependent moduli factors, and determining the corresponding structure constants. Such applications could further clarify how the intermediate spectra depend on the underlying model.

\appendix

\section{Zamolodchikov's recursion relation and cancellation of residues}\label{appendix:residue-cancellation}
In this appendix, we present the computation of the cancellation of residues in Section \ref{subsec:disk-2-pt-fct}. The computation follows closely Section 3 in \cite{Ribault2019InterpolateMinimalModels}.

The integral we consider is:
\begin{equation*}
    \int \dd P \frac{\sin(2\pi b^{-1}P)\sin(2\pi bP)}{\cos(2\pi bP)-\cos(2\pi bP_{\mathrm{s}})} \hat{C}^P_{P_1,P_2} \cF_{\lambda_P}(\lambda_1,\lambda_2,\lambda_2,\lambda_1;|z|^2).
\end{equation*}
The goal is to show: for any $m,n\in\bbZ_+$, the residue of the integrand at $P=P_{(-m,n)}$ (pole of $\hat{C}^P_{P_1,P_2}$) is the negative of the residue at $P=P_{(m,n)}$ (pole of the conformal block). Note that we have
\[ \sin(2\pi b^{-1}P_{(-m,n)})=\sin(2\pi b^{-1}P_{(m,n)}), \qquad \cos(2\pi bP_{(-m,n)})=\cos(2\pi bP_{(m,n)}). \]
Thus, it suffices to show:
\begin{equation}\label{eq:residue-cancellation}
    \frac{\mathrm{Res}\left(\cF_{\lambda_P}(\lambda_1,\lambda_2,\lambda_2,\lambda_1;|z|^2) ; P=P_{(m,n)}\right)}{\cF_{\Delta_{(-m,n)}}(\lambda_1,\lambda_2,\lambda_2,\lambda_1;|z|^2)}
    = - \frac{\mathrm{Res}\left(\sin(2\pi bP)\hat{C}^P_{P_1,P_2};P=P_{(-m,n)}\right)}{\sin(2\pi bP_{(m,n)})\hat{C}^{P_{(m,n)}}_{P_1,P_2}}.
\end{equation}

For the LHS of \eqref{eq:residue-cancellation}, the conformal block has the following expression:
\[ \cF_\lambda(\lambda_1,\lambda_2,\lambda_2,\lambda_1;|z|^2) = \cF^{(0)} \rho^\lambda H_\lambda(\rho), \]
where $\cF^{(0)}$ is a $\lambda$-independent prefactor, and $\rho$ is a function of $z$ (which is 16 times the elliptic nome). By Zamolodchikov's recursion relation, the residue of $H_\lambda(\rho)$ is given by 
\[ \mathrm{Res}\left(H_\lambda(\rho);\lambda=\Delta_{(m,n)}\right) = \rho^{mn}R_{m,n} H_{\Delta_{(-m,n)}}(\rho), \]
where
\[ R_{m,n}=\frac{-2P_{(0,0)}P_{(m,n)}}{\prod_{r=1-m}^{m}\prod_{s=1-n}^{n}2P_{(r,s)}} \prod_{r\overset{2}{=}1-m}^{m-1}\prod_{s\overset{2}{=}1-n}^{n-1}\prod_{\pm}\left(P_2\pm P_1+P_{(r,s)}\right)^2. \]
Here, the notation $\overset{2}{=}$ means that the index increases in steps of $2$.
Note that the residue for $P$ at $P_{(m,n)}$ equals $\frac{1}{2P_{(m,n)}}$ times the residue for $\lambda=P^2-P_{(1,1)}^2$ at $\Delta_{(m,n)}$.
Thus, the LHS of \eqref{eq:residue-cancellation} equals:
\[ \frac{-P_{(0,0)}}{\prod_{r=1-m}^{m}\prod_{s=1-n}^{n}2P_{(r,s)}} \prod_{r\overset{2}{=}1-m}^{m-1}\prod_{s\overset{2}{=}1-n}^{n-1}\prod_{\pm}\left(P_2\pm P_1+P_{(r,s)}\right)^2. \]

For the RHS of \eqref{eq:residue-cancellation}, \cite[Eq. (38)]{Ribault2019InterpolateMinimalModels} gives:
\[ \frac{\mathrm{Res}\left(\prod_{\pm}{\Gamma_b}(b\pm2P);P=P_{(-m,n)}\right)}{\prod_{\pm}{\Gamma_b}(b\pm2P_{(m,n)})} = \frac{P_{(0,0)}}{\prod_{r=1-m}^{m}\prod_{s=1-n}^{n}2P_{(r,s)}}. \]
Note that there is a difference of factor 2, since the residue as a function of $P$ is a half of the residue as a function of $2P$.
\cite[Eq. (37)]{Ribault2019InterpolateMinimalModels} gives
\[ \frac{\prod_{\epsilon_1, \epsilon_2, \epsilon_3 = \pm} \Gamma_b\left(\frac{b + b^{-1}}{2} + \epsilon_1 P_1 + \epsilon_2 P_2 + \epsilon_3 P_{(m,n)}\right)}{\prod_{\epsilon_1, \epsilon_2, \epsilon_3 = \pm} \Gamma_b\left(\frac{b + b^{-1}}{2} + \epsilon_1 P_1 + \epsilon_2 P_2 + \epsilon_3 P_{(-m,n)}\right)} = \prod_{r\overset{2}{=}1-m}^{m-1}\prod_{s\overset{2}{=}1-n}^{n-1}\prod_{\pm}\left(P_2\pm P_1+P_{(r,s)}\right)^2. \]
Multiplying these two equations and using the following expression proves \eqref{eq:residue-cancellation}:
\begin{align*}
    \hat{C}^P_{P_1,P_2} = {}& \frac{16P \prod_{\pm} \sin(\pi b^{\pm 2}) \Gamma_b(\pm 2P)}{\prod_{\epsilon_1, \epsilon_2, \epsilon_3 = \pm} \Gamma_b\left(\frac{b + b^{-1}}{2} + \epsilon_1 P_1 + \epsilon_2 P_2 + \epsilon_3 P\right)} \\
    ={}& -\frac{4\prod_{\pm} \sin(\pi b^{\pm 2})}{\sin(2\pi bP)} \frac{\prod_{\pm}\Gamma_b(b\pm 2P)}{\prod_{\epsilon_1, \epsilon_2, \epsilon_3 = \pm} \Gamma_b\left(\frac{b + b^{-1}}{2} + \epsilon_1 P_1 + \epsilon_2 P_2 + \epsilon_3 P\right)}.
\end{align*}

\bibliographystyle{alpha}
\bibliography{cibib}

\end{document}